\documentclass[final,5p,times,twocolumn]{elsarticle}
\usepackage{amsmath,amssymb,amsfonts}
\usepackage{graphicx}
\usepackage{multicol}
\usepackage{multirow}
\usepackage{booktabs}
\usepackage[utf8]{inputenc}
\usepackage{colortbl}
\usepackage[usenames,dvipsnames]{xcolor}
\usepackage{enumitem}
\usepackage{wrapfig}
\usepackage[edges]{forest}
\usetikzlibrary{shadows.blur}
\usepackage{tabularx}
\usepackage{hyperref}
\newcommand{\blue}[1]{\textcolor{black}{#1}}

\journal{Journal of Network and Computer Applications}

\begin{document}
\begin{frontmatter}

\title{AI Augmented Edge and Fog Computing: Trends and Challenges}

\author{Shreshth Tuli\fnref{one}}
\author{Fatemeh Mirhakimi\fnref{two}}
\author{Samodha Pallewatta\fnref{three}}
\author{Syed Zawad\fnref{four}}
\author{Giuliano Casale\fnref{one}}
\author{Bahman Javadi\fnref{two}}
\author{Feng Yan\fnref{four}}
\author{Rajkumar Buyya\fnref{three}}
\author{Nicholas R. Jennings\fnref{five}}
\fntext[one]{Imperial College London, UK}
\fntext[two]{Western Sydney University, Australia}
\fntext[three]{University of Melbourne, Australia}
\fntext[four]{University of Nevada, Reno, USA}
\fntext[five]{Loughborough University, UK}

\begin{abstract}
In recent years, the landscape of computing paradigms has witnessed a gradual yet remarkable shift from monolithic computing to distributed and decentralized paradigms such as Internet of Things (IoT), Edge, Fog, Cloud, and Serverless. The frontiers of these computing technologies have been boosted by shift from manually encoded algorithms to Artificial Intelligence (AI)-driven autonomous systems for optimum and reliable management of distributed computing resources. Prior work focuses on improving existing systems using AI across a wide range of domains, such as efficient resource provisioning, application deployment, task placement, and service management. This survey reviews the evolution of data-driven AI-augmented technologies and their impact on computing systems. We demystify new techniques and draw key insights in Edge, Fog and Cloud resource management-related uses of AI methods and also look at how AI can innovate traditional applications for enhanced Quality of Service (QoS) in the presence of a continuum of resources. We present the latest trends and impact areas such as optimizing AI models that are deployed \textit{on} or \textit{for} computing systems. We layout a roadmap for future research directions in areas such as resource management for QoS optimization and service reliability. Finally, we discuss blue-sky ideas and envision this work as an anchor point for future research on AI-driven computing systems.
\end{abstract}

\end{frontmatter}
\section{Introduction}
\noindent

In the past decade, the evolution of our digital lives has accelerated across multiple facets, including efficient computation~\cite{gill2019transformative}, communication~\cite{shi2020communication} and transportation~\cite{cluteril}, making our lives simpler and more convenient. This evolution has been driven by several factors such as the rising concern for climate change and sustainable computing~\cite{tuli2021hunter}, the expected end of Moore's law for silicon-based compute systems~\cite{theis2017end} and the recent lifestyle-changing pandemics~\cite{ndiaye2020iot} to name a few. With changing user demands and application scenarios, novel techniques are required to fuel further growth for high fidelity and scalable computation. There are two trends at the center of this growth: Artificial Intelligence (AI) and the Internet of Things (IoT). In the context of resource management, the field of AI aims to build intelligent entities that automate the process of dynamically making various design decisions for industrial computational deployment. The shift from relying on hand-encoded algorithms and human domain experts to AI or Machine Learning (ML) agents has enabled computation on scale and facilitated servicing the computational needs of the rising world population~\cite{gill2019transformative}. The other field, IoT, promises ubiquitous connectivity across various computational and networking devices using the Internet. It presents a broad spectrum of computational resources, ranging from resource-limited devices at the Edge of the network~\cite{mao2017mobile} to the heavy physical or virtual machines in the Cloud~\cite{tuli2019fogbus}, all connected to the users via gateway devices. Fog computing devices connect Edge and Cloud resources, giving rise to a paradigm that holistically considers the entire continuum of resources from Edge to the Cloud, often referred to as \textit{Fog~Continuum}~\cite{bittencourt2018internet}. 
% TODO: Replace tuli2021hunter with zhu2021green

\begin{figure*}[t]
    \centering
    \includegraphics[width=0.8\linewidth]{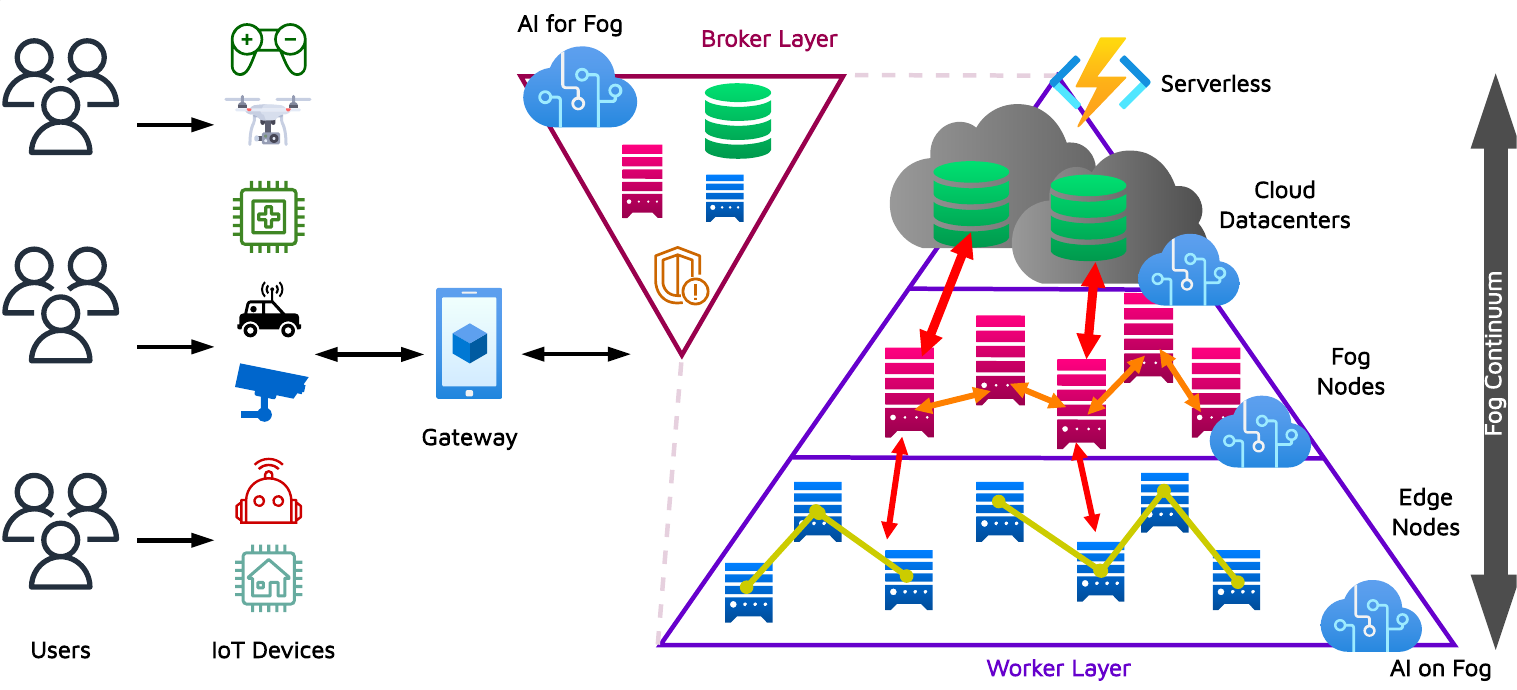}
    \caption{AI \textit{on} and \textit{for} the Fog continuum.}
    \label{fig:intro}
\end{figure*}

The convergence of AI and Fog Continuum, presents a massive opportunity for research and enterprise. Technological research organizations, such as Gartner, predict that in the coming years, AI-augmented computing will be at the forefront of technical advancements in Internet and Communication Technologies (ICT)~\cite{costello2019gartner}. When deploying AI-based applications on Fog continuum or leveraging AI to manage running applications, novel resource management issues arise corresponding to the maintenance of optimal Quality of Service (QoS). As part of this paper, we explore the latest trends in the domain of AI-augmented resource management and the challenges it presents to deliver upon the promise of improving QoS of existing and next-generation computational infrastructures.

\subsection{Motivation of Research in AI-Based Augmentation}
\begin{figure}
    \centering
    \includegraphics[width=0.95\linewidth]{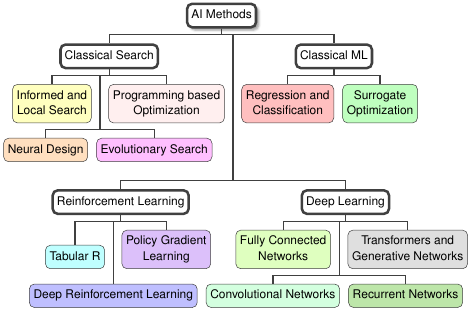}
    \caption{A brief taxonomy of AI methods for Fog systems that extends the one proposed by Russell and Norvig~\cite{aima}.}
    \label{fig:taxonomy}
\end{figure}

% What challenges can AI solve in Fog? AI4 Fog and Fog4AI. Metrics, heterogeneity at workload and hosts, unreliability in nodes.
\blue{A typical Fog environment consists of two computational layers: broker and worker (see Figure~\ref{fig:intro}). The worker layer consists of generic compute nodes that execute incoming applications by processing incoming data from the users and return the results \textit{via} gateway devices~\cite{tuli2019fogbus} (see nodes in the purple triangle in Figure~\ref{fig:intro}). The broker layer consists of compute nodes that monitor and manage the back-end infrastructure, including the worker nodes (see nodes in the inverted red triangle in Figure~\ref{fig:intro}). This includes deciding where to deploy/place incoming applications as tasks or migrate running tasks to optimize system performance.} This difference in broker and worker roles are tied closely with the classification of AI based approaches of AI \textit{on} and \textit{for} Fog that we describe later. Recent research in AI has shown some promise in the direction of improving QoS of Fog systems, thanks to their higher inference speeds and accuracy compared to classical techniques~\cite{liang2020ai}. AI research for Fog systems has spanned diverse categories including (1) classical AI that covers informed and uninformed search methods, (2) machine learning that encompasses unsupervised, supervised and semi-supervised methods, (3) reinforcement learning that includes tabular and deep reinforcement methods, and (4) deep learning that uses deep neural networks as function approximators to model complex relationships across data in Fog systems~\cite{aima,goodfellow2016deep}. \blue{A brief taxonomy from~\cite{aima} is presented in Figure~\ref{fig:taxonomy}. We shall leverage this taxonomy in Section~\ref{sec:sota} to discuss and classify state-of-the-art AI research for Fog systems.}

AI-based augmentation of Fog systems has traditionally been in two major directions. First, where AI models have replaced conventional applications, for instance Deep Neural Networks (DNNs) have replaced prior methods in domains such as traffic surveillance using computer vision, chat bots using natural language processing and smart homes using robotics~\cite{shi2020communication, lstm_vae, amini2020learning}, giving fast, scalable and accurate results. This entails augmenting the workloads that are run \textit{on} the Fog worker nodes, and hence we call this domain \textit{AI~on~Fog}.  AI on Fog has been a key driving factor many AI based practical deployments, such as self-driving cars, smart-cities and automated surveillance systems~\cite{wang2020intelligent}. Second, where the AI models are used to determine optimal workload placement, service level schedules and fault remediation steps. This augments the resource management services at the broker layer \textit{for} decision making, and hence we call this domain \textit{AI~for~Fog}. This domain has been crucial for efficient resource management for modern distributed services such as Netflix and cloud platforms~\cite{tuli2021splitplace, varghese2018next}.  We elucidate the challenges presented by each paradigm below.

\textbf{AI on Fog.} This domain is primarily concerned with the applications running on the worker layer of a Fog system. As modern applications have turned heavily dependent on AI-based models, specifically those that utilize deep learning, we observe that DNNs are becoming the backbone of many industrial tasks and activities~\cite{gill2019transformative}. As the computational capabilities of devices have improved, new deep learning models have been proposed to provide improved performance~\cite{zhu2018benchmarking, li2019edge}. \blue{Moreover, many recent DNN models have been incorporated with mobile edge computing to give low latency services with improved accuracy compared to shallow networks. Specifically in time-critical complex tasks such as image segmentation, high frame-rate gaming and traffic surveillance that require latency in the order of 10-100 milliseconds~\cite{khanna2020intelligent}.} The performance of such neural models reflects directly on the reliability of application domains like self-driving cars, healthcare and manufacturing~\cite{gill2019transformative, kraemer2017fog}. The integration of such AI models with various computational systems has led to the rise of EdgeAI services, \textit{i.e}, applications that utilize AI to process data at the edge. To provide high accuracy, neural models are becoming increasingly demanding in terms of data and compute power, resulting in many challenging problems. To accommodate these increasing demands, such massive models are often hosted as web services deployed on the public Cloud~\cite{zhang2017deep}. On the other hand, mobile edge devices in typical Fog deployments face severe limitations in terms of computational and memory resources as they rely on low power energy sources like batteries, solar, or other energy scavenging methods~\cite{mao2016dynamic}. This is not only because of the requirement of low cost, but also the need for mobility in such nodes~\cite{khanna2020intelligent}. In such systems, it is possible to handle the processing limitations of massive AI models by effective preemption and prolonged job execution. However, memory bottlenecks are much harder to solve as shown in prior work~\cite{shao2020communication}. In a practical distributed edge environment where storage spaces are typically mapped to a network-attached-media, a large virtual memory imposes high network bandwidth overheads that make performing large-scale distributed computations hard~\cite{laskaridis2020spinn}. Thus, as part of this paper, we explore various methods developed to efficiently deploy and manage AI-based applications on Fog infrastructures by possibly decomposing DNNs and running distributed training and inference~\cite{li2020federated}.

\textbf{AI for Fog.} This domain is primarily concerned with resource management level decision making at the broker layer of a Fog system. The problem of efficiently managing Fog resources is hard~\cite{tuli2021cosco}. One of the challenges we face in such a system is the heterogeneity of resources across the Edge and Cloud~\cite{li2020heterogeneity, kaur2020deep, hosseinalipour2020federated}. Another challenge in industrial settings is to deliver low latencies for time-critical applications, for instance, healthcare, robotics and smart-cities. These challenges are exacerbated by modern-day applications, wherein the workloads are highly dynamic and host machines having volatile resource capabilities. Furthermore, as applications become more demanding and privacy-sensitive, Fog devices have become more prone to breakdowns, malicious attacks and intrusions~\cite{zhang2019serious}. This entails taking recovery steps required to deal with the diverse effects of system faults, such as network packet drops, memory errors or disk failures requiring different remediation steps. So far, the industrial and research landscape of Fog resource management has been dominated by heuristics and classical optimization-based methods. Such approaches have low scheduling times and work well for general cases, but due to steady-state or stationarity assumptions, they provide poor performance in non-stationary heterogeneous environments with dynamic workloads~\cite{tuli2021cosco}. To address these challenges, various AI methods have been recently proposed that utilize adaptive schemes based on evolutionary methods and reinforcement learning. These methods adapt to changing scenarios and offer promising avenues for dynamic optimization~\cite{fox2019learning}. For accurate and scalable modeling of the Fog environment, such methods use deep learning-based local search or learning models with neural networks which approximate an objective function such as energy consumption or response time~\cite{tuli2020dynamic, liu2020collaborative, basu2019learn}. However, the most accurate AI methods typically have high decision times or resource footprints, making it hard to deploy them in budget or resource-constrained settings. Thus, as part of this paper, we also explore various advancements in AI methods for efficient resource management in Fog systems~\cite{li2020federated}.

\subsection{Our Contributions}
\label{sec:contributions}
% What challenges in integrating the two?
\blue{Our primary focus lies at the intersection of the two fields of AI and Fog, particularly for resource management decision making to optimize system performance measured using metrics like systems QoS. We review a broad range of techniques developed for optimizing QoS by efficiently deploying AI applications in Fog systems (AI on Fog), utilizing AI methods for resource management decision making (AI for Fog), or both of these together.} We partition the entire resource management domain into three scopes based on the decisions we need to optimize: deployment, scheduling and maintenance.

\begin{enumerate}[leftmargin=*]
    \item \textbf{Deployment} deals with intelligent resource provisioning and versioning of workloads on Fog infrastructures to optimize QoS~\cite{tuli2021splitplace, calheiros2014workload}.
    \item \textbf{Scheduling} deals with arranging and controlling deployed workloads on compute infrastructure for QoS efficient execution~\cite{tuli2021cosco, kadota2018scheduling, matrouk2021scheduling}.
    \item \textbf{Maintenance} aims at securing, preventing and recovering from failures the deployed and scheduled workloads in a Fog environment~\cite{du2020fault, tuli2021pregan}.
\end{enumerate}

We present a comprehensive literature review of the state-of-the-art approaches in the above three scopes. We devised a search query based on the
formulated research questions: (edge computing) OR (fog computing) OR (cloud computing) AND (AI) and (resource management) OR (scheduling) OR (provisioning) OR (fault-tolerance). We classify the reviewed methods by their essential characteristics and methodologies. We identify the future directions of AI based augmentation technologies for Fog platforms.

\begin{table*}[t]
    \centering
    \caption{A comparison of our work with existing surveys based on key parameters and domain coverage}
    \begin{tabular}{cccccccccc}
    \toprule 
    \multirow{2}{*}{Ref.} & \multicolumn{4}{c}{ Fog Continuum} & \multicolumn{2}{c}{AI Domains} & \multicolumn{3}{c}{Problem Scopes}\tabularnewline
    \cmidrule{2-10} 
    & IoT & Edge & Cloud & Serverless & AI on Fog & AI for Fog & Deployment & Scheduling & Maintenance\tabularnewline
    \midrule
    \cite{yang2019integrated} &  & \checkmark &  &  & \checkmark &  &  &  & \checkmark\tabularnewline
    \cite{wang2020convergence} &  & \checkmark &  &  & \checkmark &  & \checkmark & \checkmark & \tabularnewline
    \cite{liu2021bringing} &  & \checkmark &  &  & \checkmark &  & \checkmark &  & \tabularnewline
    \cite{murshed2021machine} & \checkmark & \checkmark &  &  & \checkmark &  & \checkmark &  & \tabularnewline
    \cite{varghese2018next} &  &  & \checkmark &  &  & \checkmark & \checkmark & \checkmark & \tabularnewline
    \cite{hasan2018fault} &  &  & \checkmark &  &  & \checkmark &  &  & \checkmark\tabularnewline
    \cite{zhong2021machine} &  & \checkmark & \checkmark &  &  & \checkmark &  & \checkmark & \tabularnewline
    \cite{singh2021fog} &  & \checkmark & \checkmark &  &  & \checkmark & \checkmark & \checkmark & \tabularnewline
    \cite{nayeri2021application} & \checkmark & \checkmark & \checkmark &  &  & \checkmark & \checkmark & \checkmark & \tabularnewline
    \cite{mampage2021holistic} &  &  &  & \checkmark &  & \checkmark & \checkmark & \checkmark & \tabularnewline
    \cite{duc2019machine} &  & \checkmark & \checkmark &  &  & \checkmark & \checkmark &  & \checkmark\tabularnewline
    \cite{deng2020edge} & \checkmark & \checkmark &  &  & \checkmark & \checkmark & \checkmark & \checkmark & \tabularnewline
    This review & \checkmark & \checkmark & \checkmark & \checkmark & \checkmark & \checkmark & \checkmark & \checkmark & \checkmark\tabularnewline
    \bottomrule
    \end{tabular}
    \label{tab:related_surveys}
\end{table*}
\subsection{Related Surveys}
% Related surveys and their Limitations and contribution
As summarized in Table~\ref{tab:related_surveys}, some previous surveys have already explored the use of AI for enhancing Fog environments; however, they do not consider the diverse use cases together and cover the complexity of the domains to a limited extent. The first four surveys cover the domain of \textit{AI~on~Fog}. Yang \textit{et al.}~\cite{yang2019integrated} introduce and review AI-based methods for data integrity, specifically utilizing Blockchain and deep learning technologies. Wang \textit{et al.}~\cite{wang2020convergence} present an exhaustive review of methods for efficient deployment and scheduling of DL-based applications on Edge infrastructures. They discuss several advancements in Edge hardware for accelerating AI training and inference. However, these reviews do not include the specific advancements in DNN models that focus on deployments in heterogeneous Edge-cloud infrastructures. Liu \textit{et al.}~\cite{liu2021bringing} present several techniques to deploy massive DNNs in Edge environments, particularly focusing on model compression and neural architecture search. Here, we also consider the recent developments in distributed split neural models. Murshed \textit{et al.}~\cite{murshed2021machine} consider distributed DNN training and inference for EdgeAI applications; here, we also cover the impact on the resource management back-ends in Edge systems. 

Furthermore, there have been some recent studies that investigate the \textit{AI~for~Fog} domain. Varghese and Buyya~\cite{varghese2018next} discuss various technological advancements in Cloud computing domain that leverage AI models for task placement and scheduling. Hasan and Goroya~\cite{hasan2018fault} summarize the research in fault-tolerant Cloud computing using AI-based methods. These works ignore the effects of merging the Cloud paradigm with Edge nodes. Zhong \textit{et al.}~\cite{zhong2021machine} discuss various methods to schedule workloads in the form of containers in Edge and Cloud environments. Similar surveys by Singh \textit{et al.}~\cite{singh2021fog} and Nayeri \textit{et al.}~\cite{nayeri2021application} describe the methods for provisioning nodes and scheduling tasks in a Fog environment. Duc \textit{et al.}~\cite{duc2019machine} discuss similar methods for reliable resource provisioning in Edge-cloud environments. Mampage \textit{et al.}~\cite{mampage2021holistic} describe resource management techniques for serverless computing environments. However, these works consider AI as black-box models and do not discuss the specific advancements in the underpinning AI techniques for QoS improvement in the context of deployment, scheduling or maintenance. Finally, Deng \textit{et al.}~\cite{deng2020edge} discuss the AI methods for and on Edge platforms, but only in the context of task allocation and AI model compression. They do not discuss the use of latest technologies such as coupled-simulation~\cite{tuli2021cosco} in solving major challenges faced when utilizing AI models for efficient resource management. Further, they restrict their descriptions to edge-only environments and do not consider the complete fog continuum.

This work builds upon the previous surveys to present a holistic view of how AI models have augmented Fog systems, particularly focusing on the overlap among \textit{AI~on~Fog} and \textit{AI~for~Fog} methods. We emphasize the diversity and complexity of QoS aware resource management schemes in the Fog continuum by categorizing the landscape into deployment, scheduling and maintenance related strategies. Unlike previous surveys, we present a classification of AI and fog methods that highlights the intersection between data-driven models and resource management distributed systems encompassing AI design, system modelling and workload-injection frameworks. Using such a holistic approach, we consolidate trends to present root-cause issues that limit the performance of AI or fog systems and share possible future directions to tackle them.

\subsection{Article Structure}

The rest of the paper is organized as follows: Section~\ref{sec:background} reviews the computing paradigms of IoT, Edge, Cloud and serverless and how Fog harnesses them. We describe the various service architectures and elucidate the main control knobs and optimization parameters. %Central to such control knobs and parameters are the underpinning frameworks and platforms in both ends of the spectrum: AI modeling and Fog design. These are discussed in Section~\ref{sec:integration}. It also presents various crucial auxiliary parameters researchers need to consider while working at the intersection of both fields. 
We discuss state-of-the-art AI methods in Section~\ref{sec:sota}. This section presents these methods in the scopes of deployment, scheduling and maintenance. We then perform a detailed trend analysis and methodological overlap in Section~\ref{sec:classes}. Such trend analysis facilitates in determining root-causes for current limitations and possible solutions as future directions as detailed in Section~\ref{sec:future_directions}. Finally, Section~\ref{sec:conclusions} concludes the survey.

%%%%%%%%%%%%%%%%% BACKGROUND %%%%%%%%%%%%%%%%% 

\section{Background}
\label{sec:background}

In this section, we present the various computing paradigms that form the Fog continuum, service architectures and parameters offered from the systems aspect for AI methods to exploit and optimize the overall QoS.

% SHORTEN
\subsection{Related Computing Paradigms}
We now describe the computing paradigms of Cloud, Edge and serverless. We mention their merits and limitations to motivate the need for a continuum of resources.

\textbf{Cloud Computing.} The Cloud computing paradigm consists of an inter-connected and virtualized pool of resources (computing, storage, network, etc.) that can be dynamically provisioned on-demand, as per user specifications and with minimal management effort~\cite{buyya2009cloud} (see top tier in Figure~\ref{fig:intro}). Cloud resources may be publicly accessible or privately deployed. Traditionally, workloads are run in Cloud nodes as distinct virtual machines (VMs), allowing Cloud providers to migrate running workloads from one Cloud node to another for load balancing and tuning various QoS parameters. A significant challenge in the Cloud paradigm is that Cloud datacenters are located multiple hops away from the IoT devices, which increases the data transmission time between the devices and the Cloud instances hosting the applications. To overcome these limitations of Cloud computing, a new paradigm called Edge computing was introduced to meet the service requirements of large-scale IoT applications.

\textbf{Edge Computing.} Recently, Edge computing~\cite{satyanarayanan2017emergence} has grown dramatically. The network edge, defined as the computational layer that resides closest to the end-user, is where most data sources are present. Edge computing follows the data gravity principle, \textit{i.e.}, it moves the computational resources close to the data sources or the network edge (see bottom tier in Figure~\ref{fig:intro}). This leads to a multitude of benefits~\cite{hu2017survey}. First, it offers low response times, possibly in milliseconds, crucial for time-critical tasks such as flight control, healthcare, autonomous cars and gaming~\cite{li2019edge, gill2019transformative}. Second, it allows us to build reliable systems where service resilience is provided at the node level, allowing other compute devices to act as backups and ameliorating performance degradation by reducing service downtimes using failover and fallback mechanisms~\cite{bagchi2019dependability}. A major challenge at the Edge is that devices have limited computational capabilities and therefore suffering significantly under stress. There also exists a vast amount of devices in an IoT system, giving rise to bandwidth contentions~\cite{belcastro2021evaluation}.

\textbf{Serverless Computing.} Serverless computing emerged as a solution for the complexity of Cloud and Edge computing that hides server usage and runs user codes on-demand automatically with high scalability at the function level, such that the users are only billed for the code execution time~\cite{castro2019rise}. It is agnostic to the specific set of resources we utilize, Edge or Cloud. Platforms and architectures have been recently proposed in the literature to extend serverless capabilities to Edge computing~\cite{javadi2020serverless, cicconetti2020decentralized}. In serverless, the applications use precisely the amount of resources needed at any one point in time and charge accordingly, making the costs proportional to the exact resource usage~\cite{hendrickson2016serverless}. Even though the tight integration in serverless makes it friendly for user, it also makes it hard for developers to optimize QoS when running serverless applications due to the lack of data management in serverless. Unlike containers and VMs that allow independent monitoring of each running service, serverless frameworks abstract out the active functions in the system, reducing the viability of tuning them for performance optimization.

\subsection{Shift to Fog Continuum}
There are typically several resource-constrained edge nodes in close proximity to the users and resource-abundant cloud nodes that are at multi-hop distance. This imposes the challenge of managing the resource-latency trade-off between edge and cloud layers, which fog continuum aims to address.
% Considering that there are as many as billions of Edge nodes in close proximity to the users, with only thousands of Cloud nodes accessible to all, it could lead to extremely high capital expenditure to execute services at the Edge for users across the globe. Thus, industries typically execute workloads corresponding to local users in Edge devices, whereas perform location-agnostic operations in the Cloud. 
%To put this in perspective, video streaming companies use Fog computing to possibly cache and process a popular video locally, which is then streamed to a nearby user leading to a great viewing experience. For less popular videos, they may be stored and processed in the Cloud, leading to higher buffer times, and in some cases, poorer video quality. However, Cloud resources are much more powerful, and thus, allow computation of heavy workloads that cannot be processed at the Edge, such as video recommendation systems and other time-insensitive tasks~\cite{tuli2021start, deng2021fogbus2}. 
None of the previously mentioned paradigms are ideal for building a generic computational platform for the end-users. The high latency of Cloud nodes, the unreliability of Edge devices, and the limited exposure of the resource management level controls offered by serverless frameworks motivate researchers to leverage all these paradigms in tandem, giving rise to the Fog continuum.

\textbf{Fog Continuum.} Fog is a parallel and distributed computing paradigm introduced by CISCO in 2012 as an interface between the Cloud and Edge computing systems to support latency-critical and resource-hungry application services by providing an interface between the computation and storage offered by Cloud and Edge~\cite{bonomi2012fog}. Fog introduces a hierarchical architecture with an intermediate layer between end-users and Cloud datacenters which utilizes computational, storage, and networking resources that reside within the path connecting users to the Cloud \cite{mahmud2018fog}. These resources known as Fog nodes include gateways, switches, routers, nano datacenters, Cloudlets, etc. Unlike traditional fog or mist platforms, \textit{fog continuum} is an umbrella term that includes edge only, cloud only and hybrid edge-fog-cloud resources. As Fog resources are distributed, heterogeneous, and resource-constrained compared to Cloud data centers, efficient resource provisioning and application placement algorithms are vital for harvesting the full potential of the Fog continuum.

\subsection{Services}

We now describe the various architectures utilized by the Fog continuum to service user requests. Each service architecture imposes disparate set of constraints on and the control surface expose to the underlying resource management techniques, possibly utilizing AI models.

\textbf{Infrastructure-as-a-Service (IaaS).} IaaS provides physical or virtual hardware resources (\textit{i.e.}, compute, storage, network infrastructure, etc.) on a pay-for-what-you-use basis. This eliminates the need for the initial investment in hardware and provides users with an easy and convenient way to remotely access, monitor, and configure infrastructure as a service~\cite{soualhia2019infrastructure, gill2019transformative}. IaaS gives AI-based resource managers control over provisioning, scaling of hardware resources, and deploying software on available hardware resources to maintain required levels of QoS for their deployed applications without having the responsibility of managing and controlling the underlying infrastructure. 

\textbf{Platform-as-a-Service (PaaS).} PaaS provides consumers with a development and execution environment that consists of a set of tools to create and deploy their own applications~\cite{varghese2018next, buyya2009cloud, zhang2015cloud}. This service simplifies application deployment by providing only platform level controls and hiding infrastructure level controls from the user. However, PaaS allows underpinning AI-based resource management solutions to control the applications and configurations of the platform that hosts the applications. A specific type of PaaS, Machine Learning-as-a-Service (MLaaS), presents ML technologies %are usually highly resource-intensive and the more advanced types 
such as Deep Learning require large-scale computation power to be viable. 
%Cloud or Edge systems can efficiently provide resources at scale. However, deploying such ML systems on the Cloud is challenging as it requires expertise on both ML and Cloud technologies. 
MLaaS abstracts out the deployment aspects and is used to describe Fog systems that provide out-of-the-box support for enabling ML technologies such as data pre-processing, model training and inference. Such systems aim to provide ease of use to users who are looking to develop and deploy their own machine learning applications efficiently.

\textbf{Software-as-a-Service (SaaS).} SaaS provides the highest level of abstraction by providing consumers with the capability to use applications running within Fog or Cloud resources that the service provider manages~\cite{gill2019transformative, buyya2009cloud, varghese2018next}. This only provides AI resource managers with limited capability to control certain application configurations. This is because the underlying architecture and application capabilities are controlled and managed by the service provider.

% \textbf{Artificial Intelligence-as-a-Service (AIaaS).} Although ML is a subset of AI, the set of functionalities presented by AIaaS may largely differ from those of MLaaS. While AIaaS can include a service for any task that should be done \textit{intelligently}, it often includes rule-based process automation that utilizes AI models other than classical and modern ML techniques. These include search strategies like local and evolutionary search schemes for optimization~\cite{aco, impso, zhu2016pso}, decision making systems like Multi-Armed bandits (MABs)~\cite{tuli2021splitplace}, constraint satisfaction solvers~\cite{liu2020pruning}. AIaaS is an umbrella term that is only considered an MLaaS once the process relies on some form of learning, such as supervised, unsupervised or semi-supervised. Supervised learning uses labelled data for training which is not used in unsupervised learning. Semi-supervised learning uses a combination of labelled and unlabelled data. For resource management solutions, this service model presents the same control parameters as MLaaS, with non data-driven functions such as search schemes.

\subsection{Optimization Parameters}

We now describe the various Quality of Service (QoS) parameters of a Fog system that we expect an AI-based resource manager to optimize for ideal system performance.

% \textbf{Quality of Service (QoS).} QoS defines the expected performance of an application/service using quantitative metrics such as response time, energy, and cost. In Cloud and Edge environments, consumers and providers negotiate these QoS parameters to establish a Service Level Agreement (SLA)~\cite{buyya2009cloud}. SLAs are critical in deadline-oriented tasks such as flight management systems, self-driving car networks and gaming. As IoT applications are heterogeneous in their characteristics (e.g., time-sensitive healthcare applications, data-intensive surveillance applications, etc.), QoS-aware scheduling mechanisms are necessary to utilize resource-constrained devices.

\textbf{Response Time.} This parameter indicates the service delivery time. Within distributed Fog environments, the response time of a service depends on multiple parameters such as data transmission time, propagation time, processing time, and service deployment time \cite{mahmud2020application}.  Thus, Fog resource provisioning and application scheduling consider the response time as a vital parameter for utilizing distributed and heterogeneous Fog resources, along with remote Cloud datacenters to prioritize applications/services with stringent latency requirements for placement within Fog environments. In Cloud and Edge environments, consumers and providers negotiate these QoS parameters to establish a Service Level Agreement (SLA)~\cite{buyya2009cloud}. SLAs are critical in deadline-oriented tasks such as flight management systems, self-driving car networks and gaming. As IoT applications are heterogeneous in their characteristics (e.g., time-sensitive healthcare applications, data-intensive surveillance applications, etc.), QoS-aware scheduling mechanisms are necessary to utilize resource-constrained devices.

\textbf{Cost.} The cost of using Cloud and Edge environments depends on the type of service used by the consumer and the pricing model (\textit{i.e.}, on-demand, reserved or spot pricing) employed by the service provider. Cloud allows potential cost savings in case of computation on large-scale. Due to the limited computation capacity of the Fog nodes, novel pricing models are introduced for Fog environments~\cite{mahmud2020profit}. Thus, Fog application placement aims to reach a tradeoff between cost and response time, to minimize the cost of deployment while satisfying deadline requirements of the applications~\cite{deng2020optimal}.

\textbf{Energy.} IoT is highly scalable, with a large number of sensors generating a significant amount of data for processing. This results in higher energy consumption and carbon footprint in Cloud datacenters during data transmitting and processing~\cite{oma2018energy}. Fog continuum, with its distributed architecture, has the potential to achieve higher energy efficiency by relying on low-power edge nodes when possible~\cite{gill2019transformative, mahmud2020application}, but is limited by the energy and computation capacity of the Fog nodes~\cite{mahmud2020application}. This motivates resource provisioning and application placement algorithms to reach a tradeoff between time and energy in an IaaS or PaaS platform~\cite{ghanavati2020energy}. When utilizing resource-heavy AI models for resource provisioning in broker nodes, the brokers themselves can lead to high energy consumption. This makes it crucial to develop AI methods that are energy efficient also in terms of their inference.

\textbf{Reliability.} Reliability of Fog systems is quantitatively defined as a probability measure of how frequently a system delivers the services it has been designed for. Edge and Fog nodes are prone to different types of failures, including hardware failures, software failures, network failures and resource overflow~\cite{bagchi2019dependability}. Dynamic issues such as battery constraints, connection fluctuations, resource availability, and mobility problems can contribute to the complexity of the reliability of such systems~\cite{carvalho2021edge}. These failures are likely to be more frequent in Edge and Fog servers due to their geographical dispersion, distributed deployment, and lack of maintenance and support from providers. Even a small failure probability per node is cascaded by the presence of a large number of interconnected nodes. Therefore, reliable Fog systems must be implemented that has a low failure rate, and when it does fail, it recovers quickly.

\textbf{Accuracy.} We use accuracy as a general term to highlight the performance of an AIaaS/MLaaS service in terms of the closeness of model output to the true or expected outputs. This can include classification performance, detection accuracy or prediction error. Several metrics exist in literature to measure the performance of an AI model, such as precision, recall, F1 score, confusion matrix and area under the receiver operating characteristic (AUROC). When deploying AI-based workloads on Fog systems, it is crucial that the choice of AI models is based on the accuracy specifications from the user. Some application use-cases, such as healthcare, require extremely accurate results. On the other hand, other scenarios, for instance autonomous systems, need near real-time inference. AI models have distinct performance and inference times.

\subsection{Synergy with Industrial IoT / Industry 5.0 Applications}

\blue{Growth in adoption of various technologies including  Industrial Internet of Things (IIoT), Industry 5.0 and aforementioned computing systems has been unprecedented in recent years and as a result several industries are utilizing these technologies to improve their productivity and services~\cite{liu2017edge}. We now provide a brief overview for some important industrial applications under the umbrella terms of Industry 5.0 and how they relate to performance parameters including response time, cost, energy, accuracy and reliability when they adopt Cloud, Fog and Edge platforms. We consider energy as an indicator of the carbon footprint of the different services.} The overview of this analysis is presented in Table~\ref{tab:industry} where the importance level of parameters are classify as high, medium, and low.

\begin{table}[t]
      \caption{The importance level of performance parameters in different industries.}
        \label{tab:industry}
    \resizebox{\linewidth}{!}{
    \begin{tabular}{@{}lccccc@{}}
    \toprule 
    Industry & Resp. Time & Cost & Energy & Accuracy & Reliability\tabularnewline
    \midrule
    Agriculture & Medium & High & Medium & Medium & High\tabularnewline
    Healthcare & High & Medium & Medium & High & High\tabularnewline
    Construction & Low & High & Low & Medium & Low\tabularnewline
    Food & Medium & High & Medium & Medium & Medium\tabularnewline
    Transport & High & Low & Medium & High & High\tabularnewline
    Textile & Low & High & Medium & Medium & High\tabularnewline
    Gaming & High & Low & Low & Medium & Medium\tabularnewline
    Aviation & High & Low & High & High & High\tabularnewline
    Smart Cities & Medium & Medium & High & High & High\tabularnewline
    \bottomrule \vspace{-12pt}
    \end{tabular}}
\end{table}

\begin{figure}[t]
    \centering
    \includegraphics[width=0.8\linewidth]{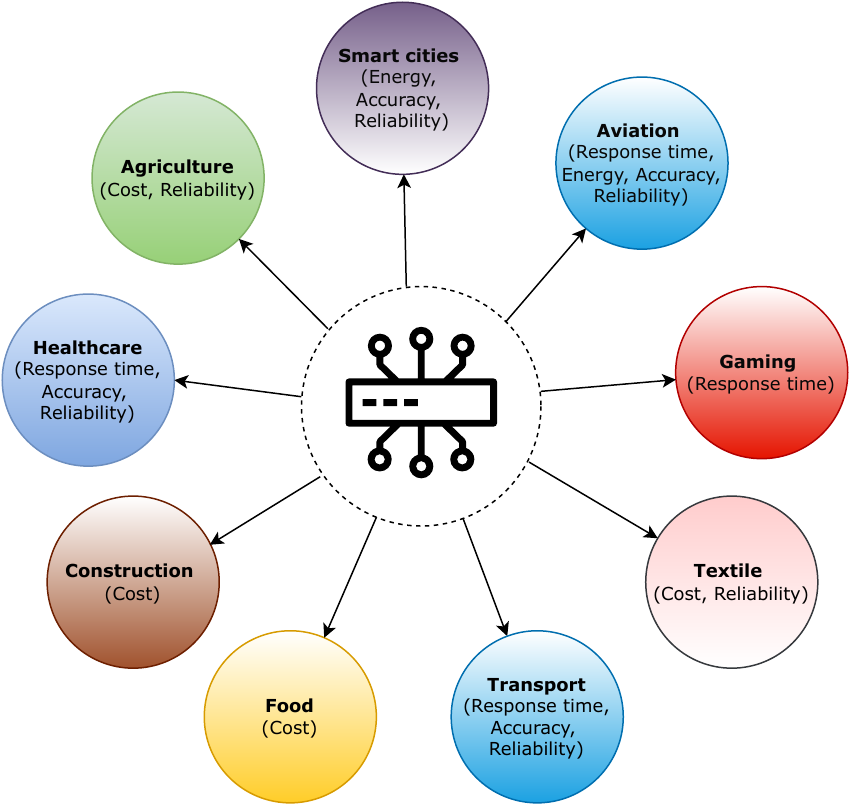}
    \caption{Critical performance metrics of Fog continuum in Industry 5.0 Applications.}
    \label{fig:applications}
\end{figure}

\blue{The \textit{Agriculture} industry widely uses various sensors for monitoring humidity, temperature, soil moisture to better control and maintain the plants and trees in the large scale agricultural fields~\cite{misra2020iot}. Important performance parameters for Fog systems would be cost and reliability as they have direct impact on the final cost and the quality of the harvest. System response time, accuracy and energy are in the medium level of importance. \textit{Healthcare} leads to the adoption of various sensors for patient monitoring and providing real-time feedback to the patient and caregivers~\cite{kumari2018fog}. Healthcare systems need low response time with high accuracy and reliability as they need to provide real-time response. The \textit{Construction} industry aims to keep tracking of the projects, and site safety. The most important metric is cost which is the main decision point for the adoption of such systems in the construction industry~\cite{abioye2021artificial}. \textit{Food} industry has widely adopted IIoT, Cloud and AI in different stages including production, transport, storage and consumption which led to the proposing of ``Internet of Food''~\cite{boulos2015towards}. The potential Fog system for this industry should be very cost effective to minimize the overall cost of the food. The \textit{Transport} industry aim to make travel more efficient by utilizing a large number of IIoT sensors, especially by the advent of autonomous vehicles~\cite{nikitas2020artificial}. Here, reliability, accuracy and response time are the most important metrics for self-driving cars. The \textit{Textile} industry uses Fog continuum in smart textile for cost effective and reliable system to curtail supply chain costs. The \textit{Gaming} industry is the growing entertainment industries, the quality of user experience is highly dependent on low latency and reliable response to the users. The \textit{Aviation} industry is now leading to the new evolutionary era called Aviation 5.0 impacting manufacturing, through aircraft operation and air traffic management. Reliability is the key importance metrics for such a system. This development of \textit{Smart cities} spans from intelligent traffic management to trash collection and air quality control. The main performance metrics for such a system are energy, accuracy and reliability.  
}
An overview of highly important performance metrics for Fog continuum systems adopted in industrial applications is illustrated in Fig.~\ref{fig:applications}, which indicated reliability is the most common metrics in these applications. The rest of the discussion considers all mentioned metrics used to measure the performance of AI based resource management solutions. However, the specific choice of metrics is subject to the application use-case and deployment scenario as mentioned above.

%%%%%%%%%%%%%%%%% AI INTEGRATION IN SYSTEMS %%%%%%%%%%%%%%%%% 

\section{AI Integration in Systems}
\label{sec:integration}

Considering the background discussion in Section~\ref{sec:background}, we have established the control surface provided by Edge and Cloud paradigms for resource management. We also present the parameters optimized by AI models to generate management decisions in Fog systems. This needs extensive integration between the Fog systems and AI methods. To this end, a plethora of approaches have been developed, both at simulation and deployment levels, which provide an interface between the two technologies. We discuss these interfacing technologies in this section.

\subsection{Simulators and Frameworks for Fog Research}

We first discuss the tools that allow modeling and testing of Fog systems.

\subsubsection{Simulated Platforms}

A simulated platform enables researchers to test their methods at scale quickly. However, as simulators are approximations of the physical systems, they may provide noisy results or deviate from observations.

Popular Fog simulators, such as \textsf{iFogSim}, provide a modular, event-driven simulation platform, created on top of CloudSim, a widely used simulator for Cloud environment simulations~\cite{gupta2017ifogsim, calheiros2011cloudsim}. iFogSim enables simulation of distributed and heterogeneous Fog nodes and scheduling of IoT based application workflows. Prior work~\cite{shahidinejad2021metaheuristic,suryadevara2021energy,etemadi2021cost, tuli2020dynamic} uses this simulator to analyze a wide range of scheduling algorithms such as evolutionary algorithms, machine learning, deep learning and reinforcement learning algorithms.  Another CloudSim based simulator, \textsf{IoTSim-Edge}, allows users to test IoT infrastructure and framework by providing a testbed for deploying IoT Edge devices as a simulation in a single application~\cite{jha2020iotsim}. IoTSim-Edge also separates the broker and worker layers by explicitly defining an \textit{Edge Broker} that acts as a simulated Fog device that manages Edge resources, and an \textit{Edge Device} as simulated worker nodes.  Similarly, \textsf{PureEdgeSim}~\cite{mechalikh2019pureedgesim} takes an edge focused control of Fog systems, particularly used for disease diagnosis~\cite{javaid2021diagnose} and fuzzy tree based decision making~\cite{mechalikh2019scalable}. Others, like \textsf{SimEdgeIntel}, provides cross-platform and cross-language support, thus enabling easy integration of machine learning-based resource management policies~\cite{wang2021simedgeintel}. It supports mobility modeling, network configuration and implementation of multiple handover mechanisms. A similar simulator, \textsf{Deep FogSim}, is designed to support large-scale evaluations of the delay-energy performance of Conditional Neural Networks (CDNNs) within Fog environments~\cite{scarpiniti2021deepfogsim}. It provides a software platform to model computing and network aspects of Fog environments and simulates the performance of the inference phase of CDNNs on top of Edge or Cloud nodes.

\textsf{ECSim++} is a simulator~\cite{nguyen2018ecsim++} that extends the OMNetpp++~\cite{varga2010omnet++}, which presents capabilities of power control and cache management, making it more realistic than other simulated devices. \textsf{RelIoT} This is a reliability simulator for IoT-based Fog systems~\cite{ergun2020reliot} and presents metrics such as power consumption, execution time, breakdown time and network characteristics such as throughput, delay, network and jitter. Unlike other simulators, it offers several combinations of reliability metrics to measure the fault resilience of a Fog system. \textsf{Yet Another Fog Simulator (YAFS)}~\cite{lera2019yafs} is a simulator that allows users to monitor network topologies, device resources and network resources. Unlike other simulators, it includes network path routing and user or device level movement as part of the control knobs it offers. A serverless simulator, \textsf{SimFaaS}~\cite{mahmoudi2021simfaas} acts as a platform with serverless functionalities. It contains out-of-the-box support for simulating essential serverless properties such as \textit{cold/warm starts} and auto-scaling. It supports the stateless/function based programming paradigm and has been demonstrated to effectively simulate real usage scenarios~\cite{mahmoudi2021simfaas}. However, it still lacks support for simulating heterogeneous systems, node failures and large-scale deployments. 

Apart from the above simulators, there are also simulators such as \textsf{EmuFog}~\cite{mayer2017emufog}, \textsf{FogTorch}~\cite{brogi2017qos}, \textsf{BigHouse}~\cite{meisner2012bighouse} and \textsf{Sim4DEL}~\cite{liu2021discrete}. They are all simulators that focus on other aspects of Cloud systems such Fog topologies, storage and sensor infrastructures, accurate device simulations, streaming systems and federated deep Edge learning. There are also two Cloud-based Fog and Edge device simulators: \textsf{Azure IoT}~\cite{stackowiak2019azure} and \textsf{AWS IoT device simulator}~\cite{awsiotsimulator}. These focus on simulating large-scale IoT systems with support for simulating thousands of devices, serverless functions within Cloud VMs and integrating live sensors and actuators. As such, these two can be used since they provide support using their large pool of back-end Cloud resources. 

\subsubsection{Physical Platforms}
For credible AI-augmented Fog research, testing developed solutions on emulators that duplicate industrial deployment scenarios on physical platforms is increasingly important.

\textsf{OpenStack} is an open-source platform developed by Rackspace Inc. and NASA, originally developed for Cloud environment, but later also extended to support Edge devices, thanks to its modularized APIs~\cite{sefraoui2012openstack}. OpenStack has custom hypervisor drivers that can support a variety of virtualization technologies such as KVM, QEMU, UML, Xen, VMware, Docker and many more, making it very versatile option for Edge virtualization. Other platforms, such as \textsf{KubeEdge}~\cite{wang2020kubeedge} and \textsf{OpenEdge}~\cite{openedge}, are based on Kubernetes virtualization technology~\cite{kristiani2018implementation}. They provide functionalities for efficient communication between Edge and Cloud as well deployment of various AI-based applications~\cite{wang2020kubeedge}.  They also contain APIs that control assignment of device resources to different workloads, which allows for efficient use of resources for the already resource-constrained Edge devices.

Another framework, \textsf{FogBus}, facilitates IoT-Fog-Cloud integration to run multiple applications using platform-independent interfaces provided by the platform~\cite{tuli2019fogbus} by following master-worker topology where master nodes known as Fog Brokers are responsible for delegating data processing tasks to the worker Fog nodes. Similarly, \textsf{EiF} \textit{i.e.}, Elastic Intelligent Fog,~\cite{an2019eif} is a framework that supports AI-based service migration, predictive network resource allocation and predictive QoS-aware orchestration along with support for distributed AI. A recent framework, \textsf{COSCO}, \textit{i.e.}, Co-Simulation based Container Orchestration (COSCO)~\cite{tuli2021cosco}, is a  framework that presents AI-based resource management modules to not only utilize the workload resource utilization characteristics, but also simulated characteristics at a future state of the system. The interleaved execution of AI models and coupled simulation (referred to as co-simulation in literature) enables long-term optimization~\cite{tuli2021mcds} and quick adaptation in volatile system settings~\cite{tuli2021gosh, tuli2021pregan}.

\subsection{AI Benchmarks for Fog Systems}
\begin{table}[t]
    \centering
    \caption{Comparison of AI based benchmarks for Fog systems in terms of workload coverage.}
    \resizebox{\linewidth}{!}{
    \begin{tabular}{@{}lccccccc@{}}
    \toprule 
    \multirow{2}{*}{Benchmark} & AI & AI & \multirow{2}{*}{Robotics} & Vision /  & NLP / & Trans- & \multirow{2}{*}{GANs}\tabularnewline
    & Search & Filtering &  & CNN & RNNs & formers & \tabularnewline
    \midrule 
    DeFog \cite{mcchesney2019defog} & \checkmark &  &  & \checkmark &  &  & \tabularnewline
    AIoTBench \cite{luo2018aiot} &  &  &  & \checkmark &  &  & \tabularnewline
    AIBench \cite{gao2018aibench} &  & \checkmark &  & \checkmark & \checkmark & \checkmark & \checkmark\tabularnewline
    EdgeAIBench \cite{hao2018edgeaibench} &  &  &  & \checkmark & \checkmark &  & \tabularnewline
    EdgeBench \cite{das2018edgebench} & \checkmark &  &  & \checkmark &  &  & \tabularnewline
    IoTBench \cite{iotbench} &  &  & \checkmark & \checkmark &  &  & \tabularnewline
    \bottomrule
    \end{tabular}}
    \label{tab:benchmarks}
\end{table}

For research related to \textit{AI on Fog}, several workload like benchmarks have been utilized to test the efficacy of Fog systems, such as Raspberry Pi clusters, when dealing with AI-based applications. These are summarized in Table~\ref{tab:benchmarks}.

A popular Fog benchmark, \textsf{DeFog}~\cite{mcchesney2019defog}\footnote{DeFog: \url{https://github.com/qub-blesson/DeFog}.}, consists of six real-time heterogeneous workloads: Yolo, Pocketspinx, Aeneas, FogLamp, iPokeMon and RealFD. Yolo uses Convolution Neural Network (CNN) for object classification in images. Pocketsphinx is a Natural Language Processing (NLP) based speech-to-text synthesis engine that utilizes an AI-based search strategy. Aeneas is a text and audio synchronization tool that utilizes text to speech tools with AI-based search for minimizing speech deviation metrics. iPokeMon is an adaptation of the game Pokemon Go with simulated players and service requests for network testing in Fog. FogLamp is an application that uses aggregated sensor data and simulated data retrieval requests to test the storage bandwidth of Fog devices. RealFD uses computer vision for face detection in video streams. Other benchmarks, such as \textsf{AIoTBench}~\cite{luo2018aiot}\footnote{AIoTBench: \url{https://www.benchcouncil.org/aibench/aiotbench/index.html}.} and \textsf{EdgeAIBench}~\cite{hao2018edgeaibench}\footnote{EdgeAIBench: \url{https://www.benchcouncil.org/aibench/edge-aibench/index.html}.}, are AI-based Edge computing based benchmark suites that consist of various real-world computer vision application instances. The former consists of CNN neural networks for image classification. These include three typical heavy-weight networks: ResNet18, ResNet34, ResNext32x4d, as well as four light-weight networks: SqueezeNet, GoogleNet, MobileNetV2, MnasNet. The latter includes applications such as ICU patient monitoring and heart failure prediction using attention-based LSTMs, surveillance camera video face-detection using CNN networks and road-sign detection for autonomous vehicles using CNNs.

Another AI benchmarking suite, \textsf{AIBench}~\cite{gao2018aibench}\footnote{AIBench: \url{http://www.benchcouncil.org/AIBench/index}.}, includes a wide variety of AI applications including image classifications using CNNs, image generation using Generative Adversarial Networks (GANs), text-to-text translation using recurrent neural networks (RNNs), speech-to-text using Gated Recurrent Units (GRUs) and LSTMs, recommendation system using collaborative filtering and spatial image transformations using Transformer neural networks. \textsf{EdgeBench}~\cite{das2018edgebench}\footnote{EdgeBench: \url{https://github.com/akaanirban/edgebench}.} includes audio to text translation using AI search and object recognition using CNNs. \textsf{IoTBench}~\cite{iotbench} consists of multiple AI models run simultaneously under the same input workloads. It includes applications for image classification using CNNs and robotics workloads related to Simultaneous Localization and Mapping (SLAM) of robot environments.

Apart from the above mentioned benchmarks, several execution traces are used by state-of-the-art AI augmentation techniques as datasets for simulation based testing. \textsf{Bitbrain} consists of traces of resource utilization metrics from 1750 VMs running on BitBrain distributed datacenter~\cite{bitbraindataset}. \textsf{Azure2017} and \textsf{Azure2019} are collected from Microsoft Azure public Cloud platform and are representative workload traces across thirty consecutive days~\cite{azuredataset}. \textsf{Google Cluster} is another dataset of CPU and memory utilization traces of multiple nodes in a high-performance cluster in Google Cloud Platform~\cite{reiss2011google}. \textsf{Server Machine Dataset (SMD)}~\cite{omnianomaly} is a five-week long dataset of resource utilizations of 28 machines from a compute cluster. Other request trace datasets include \textsf{HDFS}~\cite{HDFS448A56:online}, \textsf{MHealth}~\cite{UCIMachi57:online}, \textsf{PlanetLab}~\cite{beloglaz24:online}, \textsf{Wikimedia}~\cite{wikimedia}, \textsf{Bikeshare}~\cite{bikeshare}, \textsf{Shakespeare}~\cite{tinyshak46:online}, \textsf{SETI}~\cite{javadi2009mining}, \textsf{Crawdad}~\cite{piorkowski2009dataset}, \textsf{Traffic}~\cite{sivanathan2018classifying}, \textsf{T-Drive}~\cite{yuan2010t}, \textsf{Tutoring}~\cite{metevier2019offline}, \textsf{WfCommons}~\cite{coleman2021wfcommons} and \textsf{Yahoo Webscope}~\cite{yahoo_webscope}.

\subsection{AI Modelling and Engineering}

Now that we have described the various simulation and emulation platforms for Fog systems, AI toolkits and benchmarking suites, we elucidate the challenges faced while training or running inference using an AI model. Model training is a highly resource-intensive task due to the large number of parameters in modern AI and Deep Learning models, and traditionally requires the use of high-performance clusters, Graphical Processing Units (GPUs) or Tensor Processing Units (TPUs). As such, given that Edge devices usually contain limited resources, the training overheads are significant and can take significantly longer time compared to Cloud nodes. Additionally, the hardware resources for Edge devices are used by other applications in parallel to the training process, giving rise to frequent resource contentions. Furthermore, unlike traditional clusters, Cloud engineers have little to no control over the availability of the Edge devices, making training a challenging task. 

Currently, two mainstream methods of model development and training exist: Centralized and Federated Learning (FL)~\cite{lim2020federated}. In a centralized learning system, the AI model, typically a DNN, with the training dataset is kept in a single resource-intensive machine with the training framework updating the parameter updates iteratively until convergence to minimize a developer-defined loss function~\cite{lim2020federated}. In a federated learning setup, the model to be trained is sent to multiple Edge devices that contain a subset of the training data. The model is trained locally using the local hardware and the parameter updates are iteratively aggregated into a global copy of the model. Centralized learning requires a high-end system, but does not lead to high bandwidth use as in FL. Federated learning requires nodes to synchronize models iteratively that can lead to network contentions and increased wait times. However, as only the parameter updates are shared across the nodes and not the local data, it ensures data privacy.

%%%%%%%%%%%%%%%%% SOTA %%%%%%%%%%%%%%%%% 

\section{State of the Art Methods}
\label{sec:sota}

Given the optimization parameters in Fog systems, infrastructure level constraints offered by the simulator/frameworks, we now review the state-of-the-art methods for the three aspects resource management: provisioning, scheduling and maintenance to optimize the QoS metrics including response time, cost, energy, accuracy and resiliency.

\subsection{Interface Between AI and Fog}

\textbf{Data sources and inputs for AI models.} For any resource management system in the Cloud, for instance, an AI model, this paradigm provides multi-modal data sources to analyze the system. Traditionally, these include workload resource utilization traces in the form of a fraction of CPU, RAM, Disk and Network bandwidth utilization and host resource capacities in the form of instruction per second (IPS), available RAM and Disk space and parameters of the network interface~\cite{tuli2021cosco, mao2016dynamic, jalali2019dynamic, basu2019learn, tuli2021hunter}. Other parameters include gateway bandwidths, geographical location of users and Fog nodes~\cite{lera2019yafs}, communication latencies and mobility characteristics~\cite{ye2018fault}.

\textbf{Control Knobs and outputs of AI models.} We categorize the state-of-the-art approaches for AI-augmented Fog continuum as per the decisions they aim to optimize.

\begin{enumerate}[leftmargin=*]
    \item \textit{Deployment:} This deals with the initial decisions of how to efficiently execute resource-intensive AI applications on constrained Fog systems. In \textit{AI on Fog}, this concerns with appropriate methods to deploy resource-intensive AI/ML models on constrained nodes. This entails deciding the appropriate strategy to compress AI models without compromising on performance. We discuss this aspect in Section~\ref{sec:deployment_dnn}. In \textit{AI for Fog}, this concerns the efficient allocation of resources for the input workloads. This includes the provisioning of resources, \textit{i.e.}, allocation of new and deallocation of existing Fog devices. %When we consider AI deployments on only the Edge part of a Fog environment, we call this paradigm EdgeAI. In such a setup, we have volatile and time-critical workloads. To meet their harsh deadlines, it is crucial to proactively provision resources for smooth task execution. It is crucial to avoid over-provisioning to avoid high costs and resource wastage, as well as under-provisioning to prevent contentions and faults in Fog devices. Further, based on the user deadline constraints and result fidelity requirements, appropriate processing pipelines need to be chosen. 
    We discuss this aspect in Section~\ref{sec:deployement_provisioning}.
    \item \textit{Scheduling:}  This deals with scheduling the deployed workloads on existing Fog infrastructure. This concerns with optimal placement of tasks on Fog nodes to optimize system QoS. As we consider a dynamic setting, our scheduling decisions also include task migration decisions, \textit{viz}, the relocation of one or a group of tasks from one node to another, allowing the system to adapt to changes in the environments, workloads or user demands. In \textit{AI~on~Fog} setups, if the incoming workloads have tasks that impose precedence constraints, such as different layers of a neural network, we categorize these as workflows and discuss relevant methods in Section~\ref{sec:scheduling_workflows}. Schedulers for independent tasks are discussed in Section~\ref{sec:scheduling_bot}. 
    \item \textit{Maintenance:} %The increasing volumes of data, requiring immediate processing and the resource constraints at the Edge are pushing the compute resources to their limits, giving rise to high chance of resource contention and node unavailability~\cite{dastjerdi2016fog}. 
    Even though there are several methods based on redundancy that have been proposed for Cloud computing systems~\cite{sharma2016reliability}, these techniques cannot be directly applied to Fog systems due to resource limitations. This means that Fog servers may have less capability to use redundancy techniques like replication~\cite{shivakumar2015ensuring}. This leads to higher response times and SLA violations that can lead to significant financial losses~\cite{nicoletti2013cloud}. Thus, it is critical to develop a mechanism for the maintenance of Fog environments. This deals with detecting faults/anomalies in real-time, discussed in Section~\ref{sec:maintenance_pred_detect}. Also, we consider works that develop appropriate proactive or reactive recovery mechanisms to prevent service downtime. They are either related to load-balancing methods that aim at preventing faults or scaling the Fog infrastructure (see Section~\ref{sec:maintenance_remediation}). 
\end{enumerate}

Most data-driven methods achieve local optimum. However, some approaches are developed to avoid getting stuck in such local optima~\cite{loshchilov2016sgdr}, although they do not guarantee achieving global optima. We now describe how data-driven AI methods can be used to solve the deployment, scheduling and maintenance challenges in fog continuum.

We now move to the three aspects of AI augmented resource management introduced in Section~\ref{sec:contributions}. We consider works that leverage either a physical framework or a simulated environment. %We describe the various simulators and frameworks with the benchmarking workloads used in these works in Section~\ref{sec:integration}.

\begin{table*}[t]
    \centering
    \caption{Summary of state-of-the-art methods for AI augmented deployment. Color scheme as per Table~\ref{tab:trends}.}
    \label{tab:deployment}
    \resizebox{\linewidth}{!}{
\begin{tabular}{@{}|p{1.65cm}|p{1.65cm}|c|c|c|c|>{\centering}p{1.8cm}|c|c|@{}}
\hline 
Decision Type & Category & Ref. & Method & Infra. & Benchmark & Framework / Simulator & Merits & Limitations\tabularnewline
\hline 
\hline 
\multirow{15}{2cm}{DNN Deployment} & AutoML & \cite{xia2019dnntune, li2019edge, zhao2021edgeml} & \cellcolor{Fuchsia!25}Policy Gradient Learning & E & AIoTBench & F: Custom & Fast Execution & Generalizability\tabularnewline
\cline{2-9} \cline{3-9} \cline{4-9} \cline{5-9} \cline{6-9} \cline{7-9} \cline{8-9} \cline{9-9} 
 & \multirow{3}{1.55cm}{Model Pruning} & \cite{jiang2019model, cnn_slimming} & \cellcolor{yellow!25}Search & E+C & AIoTBench & F: Custom & Cost Efficient & Accuracy\tabularnewline
\cline{3-9} \cline{4-9} \cline{5-9} \cline{6-9} \cline{7-9} \cline{8-9} \cline{9-9} 
 &  & \cite{shao2020bottlenet++, yu2020easiedge, Huang2020} & \cellcolor{orange!25}Neural Design & E+C & AIoTBench & - & Generalizable & Overhead\tabularnewline
\cline{3-9} \cline{4-9} \cline{5-9} \cline{6-9} \cline{7-9} \cline{8-9} \cline{9-9} 
 &  & \cite{zhou2021bbnet} & \cellcolor{yellow!25}Search & E+C & AIoTBench & F: Custom & Cost Efficient & Overhead\tabularnewline
\cline{2-9} \cline{3-9} \cline{4-9} \cline{5-9} \cline{6-9} \cline{7-9} \cline{8-9} \cline{9-9} 
 & \multirow{2}{1.55cm}{Gradient Pruning} & \cite{sattler2019robust, luo2021novel} & \cellcolor{yellow!25}Search & E & AIBench & F: Custom & Low Overhead & Accuracy\tabularnewline
\cline{3-9} \cline{4-9} \cline{5-9} \cline{6-9} \cline{7-9} \cline{8-9} \cline{9-9} 
 &  & \cite{reisizadeh2020fedpaq, hamer2020fedboost, tran2019federated} & \cellcolor{orange!25}Neural Design & E+C & Shakespeare & S:Custom & Low Overhead & Accuracy\tabularnewline
\cline{2-9} \cline{3-9} \cline{4-9} \cline{5-9} \cline{6-9} \cline{7-9} \cline{8-9} \cline{9-9} 
 & \multirow{2}{1.55cm}{Low Precision} & \cite{coelho2021automatic, jain2018compensated} & \cellcolor{orange!25}Neural Design & E+C & - & - & Generalizable & Scalability\tabularnewline
\cline{3-9} \cline{4-9} \cline{5-9} \cline{6-9} \cline{7-9} \cline{8-9} \cline{9-9} 
 &  & \cite{lane2016deepx, imani2019floatpim} & \cellcolor{yellow!25}Search & E+C & AIBench{*} & F: Custom & Memory Efficient & Generalizability\tabularnewline
\cline{2-9} \cline{3-9} \cline{4-9} \cline{5-9} \cline{6-9} \cline{7-9} \cline{8-9} \cline{9-9} 
 & \multirow{4}{1.5cm}{Layer Splitting \& Early Exits} & \cite{kang2017neurosurgeon, zhang2021deepslicing} & \cellcolor{yellow!25}Search & E+C & AIBench{*} & S: Bighouse & Energy Efficient & Generalizability\tabularnewline
\cline{3-9} \cline{4-9} \cline{5-9} \cline{6-9} \cline{7-9} \cline{8-9} \cline{9-9} 
 &  & \cite{callegaro2020optimal} & \cellcolor{pink!25}Linear Programming & E & - & - & Fast Execution & Scalability\tabularnewline
\cline{3-9} \cline{4-9} \cline{5-9} \cline{6-9} \cline{7-9} \cline{8-9} \cline{9-9} 
 &  & \cite{matsubara2019distilled, teerapittayanon2017distributed, goli2020migrating} & \cellcolor{orange!25}Neural Design & E+C & AIBench{*} & - & Low Overhead & Accuracy\tabularnewline
\cline{3-9} \cline{4-9} \cline{5-9} \cline{6-9} \cline{7-9} \cline{8-9} \cline{9-9} 
 &  & \cite{gillis} & \cellcolor{purple!25}Policy Gradient Learning & C & AIBench{*} & F: AWS IoT & Memory Efficient & Generalizability\tabularnewline
\cline{2-9} \cline{3-9} \cline{4-9} \cline{5-9} \cline{6-9} \cline{7-9} \cline{8-9} \cline{9-9} 
 & \multirow{3}{2cm}{Splitting} & \cite{kaplunovich2020automatic} & \cellcolor{yellow!25}Search & C & AIBench{*} & F: AWS IoT & Generalizable & Overhead\tabularnewline
\cline{3-9} \cline{4-9} \cline{5-9} \cline{6-9} \cline{7-9} \cline{8-9} \cline{9-9} 
 &  & \cite{chen2018thriftyedge, huang2020clio, kim2017splitnet} & \cellcolor{orange!25}Neural Design & E+C & AIoTBench{*} & F: Custom & High Accuracy & Generalizability\tabularnewline
\cline{3-9} \cline{4-9} \cline{5-9} \cline{6-9} \cline{7-9} \cline{8-9} \cline{9-9} 
 &  & \cite{tuli2021splitplace} & \cellcolor{blue!25}DQN + Multi Armed Bandits & E+C & AIoTBench{*} & F: COSCO & High QoS & Overhead\tabularnewline
\hline 
\multirow{18}{2cm}{Resource Provisioning} & \multirow{8}{2cm}{Demand Prediction} & \cite{hyndman2018forecasting} & \cellcolor{red!25}Linear Regression & E+C & - & S: Custom & Fast Execution & Accuracy\tabularnewline
\cline{3-9} \cline{4-9} \cline{5-9} \cline{6-9} \cline{7-9} \cline{8-9} \cline{9-9} 
 &  & \cite{zhu2016pso} & \cellcolor{red!25}Support Vector Regression & C & Google Cluster & S: Custom & Memory Efficient & Scalability\tabularnewline
\cline{3-9} \cline{4-9} \cline{5-9} \cline{6-9} \cline{7-9} \cline{8-9} \cline{9-9} 
 &  & \cite{chen2018resource, uahs, cahs} & \cellcolor{red!25}Gaussian Process Regression & E+C & Azure2017/19 & S: Custom & High Accuracy & Scalability\tabularnewline
\cline{3-9} \cline{4-9} \cline{5-9} \cline{6-9} \cline{7-9} \cline{8-9} \cline{9-9} 
 &  & \cite{taylor2018forecasting} & \cellcolor{red!25}Modular Regression & E+C & - & - & Scalable & Accuracy\tabularnewline
\cline{3-9} \cline{4-9} \cline{5-9} \cline{6-9} \cline{7-9} \cline{8-9} \cline{9-9} 
 &  & \cite{arima, calheiros2014workload} & \cellcolor{red!25}ARIMA & C & Wikimedia & S: Custom & High Accuracy & Scalability\tabularnewline
\cline{3-9} \cline{4-9} \cline{5-9} \cline{6-9} \cline{7-9} \cline{8-9} \cline{9-9} 
 &  & \cite{xu2020hybrid} & \cellcolor{red!25}Decision Regression Tree & E & Bikeshare & - & Fast Execution & Only Univariate\tabularnewline
\cline{3-9} \cline{4-9} \cline{5-9} \cline{6-9} \cline{7-9} \cline{8-9} \cline{9-9} 
 &  & \cite{bega2019deepcog, jeddi2019water} & \cellcolor{SeaGreen!25}CNN & E+C & Custom & - & High Accuracy & Interpretability\tabularnewline
\cline{3-9} \cline{4-9} \cline{5-9} \cline{6-9} \cline{7-9} \cline{8-9} \cline{9-9} 
 &  & \cite{lstm, yazdanian2021e2lg} & \cellcolor{OliveGreen!25}LSTM & C & Custom & S: Custom & High Accuracy & Interpretability\tabularnewline
\cline{2-9} \cline{3-9} \cline{4-9} \cline{5-9} \cline{6-9} \cline{7-9} \cline{8-9} \cline{9-9} 
 & \multirow{4}{2cm}{Decision Optimization} & \cite{aco} & \cellcolor{magenta!25}Ant Colony Optimization & E+C & Bitbrain & S: Cloudsim & High util. ratio & Scalability\tabularnewline
\cline{3-9} \cline{4-9} \cline{5-9} \cline{6-9} \cline{7-9} \cline{8-9} \cline{9-9} 
 &  & \cite{uahs, cahs} & \cellcolor{magenta!25}Bayesian Optimization & E+C & Azure2017/19 & S: Custom & High util. ratio & Scalability\tabularnewline
\cline{3-9} \cline{4-9} \cline{5-9} \cline{6-9} \cline{7-9} \cline{8-9} \cline{9-9} 
 &  & \cite{zhu2016pso, chen2020self} & \cellcolor{magenta!25}Particle Swarm Optimization & C & Google Cluster & S: Custom & Memory Efficient & Scalability\tabularnewline
\cline{3-9} \cline{4-9} \cline{5-9} \cline{6-9} \cline{7-9} \cline{8-9} \cline{9-9} 
 &  & \cite{asghari2021task} & \cellcolor{magenta!25}Genetic Algorithm & C & - & S: Cloudsim & High util. ratio & Execution Time\tabularnewline
\cline{2-9} \cline{3-9} \cline{4-9} \cline{5-9} \cline{6-9} \cline{7-9} \cline{8-9} \cline{9-9} 
 & \multirow{6}{2cm}{Hybrid Provisioning} & \cite{decisionnn} & \cellcolor{green!25}Deep Surrogate Optimization & E+C & Yahoo Webscope & S: Custom & Cost Efficient & Generalizability\tabularnewline
\cline{3-9} \cline{4-9} \cline{5-9} \cline{6-9} \cline{7-9} \cline{8-9} \cline{9-9} 
 &  & \cite{semidirect} & \cellcolor{pink!25}DNN + Dynamic Prog. & E+C & - & - & High util. ratio & Scalability\tabularnewline
\cline{3-9} \cline{4-9} \cline{5-9} \cline{6-9} \cline{7-9} \cline{8-9} \cline{9-9} 
 &  & \cite{narya} & \cellcolor{LimeGreen!25}FCN + Multi Armed Bandits & C & Custom & F: Kafka & Fast \& Scalable & Exposure Bias\tabularnewline
\cline{3-9} \cline{4-9} \cline{5-9} \cline{6-9} \cline{7-9} \cline{8-9} \cline{9-9} 
 &  & \cite{chen2019iraf} & \cellcolor{LimeGreen!25}FCN + Monte Carlo Tree Search & C & - & - & High QoS & Execution Time\tabularnewline
\cline{3-9} \cline{4-9} \cline{5-9} \cline{6-9} \cline{7-9} \cline{8-9} \cline{9-9} 
 &  & \cite{xu2020recarl, bitsakos2018derp, sami2021ai} & \cellcolor{blue!25}Deep Q Learning & E+C & Google Cluster & S: Custom & Scalable & Interpretability\tabularnewline
\cline{3-9} \cline{4-9} \cline{5-9} \cline{6-9} \cline{7-9} \cline{8-9} \cline{9-9} 
 &  & \cite{xu2020deep, chen2020deep, chen2021deep} & \cellcolor{purple!25}Policy Gradient Learning & E+C & - & S: iFogSim & Low Energy & Scalability\tabularnewline
\hline 
\end{tabular}}
\end{table*}

\subsection{AI Augmented Deployment}

A summary of all AI augmented deployment methods is presented in Table~\ref{tab:deployment}. Here, a \textit{benchmark} corresponds to the workloads used to train and test the presented methods. \textit{Infra.} column represents whether the methods utilize Edge (E) or Cloud (C) or both (E+C). Asterisk with framework/simulator means that the respective papers utilize a modified version of the base platforms.

\subsubsection{AI Augmented DNN Deployment}
\label{sec:deployment_dnn}

Running training or inference procedures for AI/ML models is computationally expensive. Given that Edge devices tend to have limited compute resources that are usually shared across multiple running applications, it is essential to develop resource-efficient training and inference mechanisms to ensure short training times and resource load. \blue{Several solutions have been proposed in the past to address this.}

\textbf{AutoML.} For instance, several AutoML (automated machine learning) techniques run search in the space of neural network models, \textit{i.e.}, NAS, to find out the optimal DNN model to execute a task in a given system~\cite{xia2019dnntune, zhou2019edge, li2019edge, zhao2021edgeml}. These methods can be run to find the optimal DNN architecture for a given set of constraints such as training or inference times, memory footprint, computational requirements, etc.

\textbf{Model Pruning.} Another direction is to take existing AI models and prune their parameters to reduce overall local training cost, for instance, PruneFL~\cite{jiang2019model}. Model pruning is a commonly used strategy in ML that intelligently cuts away parts of the model architecture without compromising the quality of model inference~\cite{shao2020bottlenet++, yu2020easiedge, Huang2020}. \blue{Compared to AutoML, by pruning out parts of the model, the computational resources required to train or run inference on the model are reduced, making it more amenable to Edge and Fog systems.} Similar model pruning works are dependent on the DNN design. For instance, Generative Optimization Networks (GON)~\cite{tuli2021generative} are generative models inspired by GANs that use two neural networks in tandem: generator and discriminator. \blue{Unlike GANs, GONs do not use the generator and create new samples only using the discriminator network. Other examples include SlimGAN~\cite{slimgan}, Gradient Origin Networks~\cite{bond2020gradient} and similar GAN slimming techniques~\cite{gan_slimming}.} Similar works exist that perform model pruning for other DNN types, for instance, CNNs~\cite{cnn_slimming}. Pruning reduces both the memory and compute requirements of models. Similarly methods like BBNet utilize multiple techniques together, such as model pruning and data compression~\cite{zhou2021bbnet}. BBNet decides the optimal pruning and compression parameters using local search-based techniques.

\textbf{Gradient Pruning.} In FL systems, all worker nodes need to send gradient updates of their models over the network, which generally translates to up to gigabytes of data depending on the size of the model. \blue{To avoid bandwidth contentions, several works discuss solutions to prune the gradient updates improving memory and network efficiency in federated setups~\cite{yang2020energy}.} Some approaches, such as Sparse Ternary Compression (STC)~\cite{sattler2019robust} and Deep Gradient Compression (DGC)~\cite{luo2021novel}, employ compression mechanisms to reduce the communication bandwidth required for distributed training or inference. Other methods, like FedPaq~\cite{reisizadeh2020fedpaq}, perform periodic aggregation and quantization to reduce communication frequency. A similar approach, CMFL~\cite{luping2019cmfl}, intelligently decides which model updates give the maximum boost in performance and only sends the top-performing gradient updates. Another work, FedBoost~\cite{hamer2020fedboost}, uses \textit{ensemble training} to boost model training efficiency and offload only a small part of the ensemble to the Edge devices with predefined intervals to reduce communication overheads.   Another work, FEDL~\cite{tran2019federated}, \blue{theoretically demonstrates the relationship between convergence rate and energy consumption of an FL system and formulates the computation and communication models as a non-convex optimization program to optimize the distribution of federated networks and outperformed other learning methods.}
 
\textbf{Low-Precision.} Energy efficiency is also one of the major concerns when designing efficient ML training algorithms due to FL parties generally being battery-powered devices. Deep learning is inherently very power-consuming due to the large amounts of computation that need to be performed. As such, there has been a relatively large amount of work in energy efficiency by discrete quantization and using low-precision hardware architectures~\cite{coelho2021automatic, gong2019mixed, jain2018compensated, langroudi2019cheetah, langroudi2019deep}. This not only reduces the computational overheads, but also gives significant gains in terms of memory and energy footprints~\cite{gong2019mixed, tuli2021generative}. The level of precision cannot be changed dynamically at test time as changing precision requires re-training the models. Thus, \blue{these decisions either need to be performed at the setup time or multiple models need to be trained of different precision levels, at the cost of higher training time, to provide control to the resource manager to tradeoff between accuracy and memory footprint. Other optimization methods, such as DeepX~\cite{lane2016deepx},} focus on the development of deep learning models on mobile devices by runtime control of the memory to reduce the layer-wise operations, such that only the most important operations use larger bytes. \blue{Further, it efficiently identifies unit blocks of a DNN architecture and allocates them to local or remote memory caches depending on the access frequency, improving memory footprint.} A similar method, FloatPIM~\cite{imani2019floatpim} provides an interface between software and hardware by using Processing in-memory (PIM) to reduce memory usage. 

\textbf{Layer Splitting and Early-Exits.} For typical DNNs, it is possible to run inference without performing operations across all layers. Methods, such as Neurosurgeon~\cite{kang2017neurosurgeon} and DeepSlicing~\cite{zhang2021deepslicing}, decide the optimal layer partitions of a neural network using grid-search at run time to maximize system QoS. Others, like SplitComp~\cite{callegaro2020optimal} model the problem of deciding the optimal splitting strategy as a Markov Process and leverage Linear Programming to converge to the optimal splitting strategy. Further, \blue{to reach the best tradeoff between model accuracy and processing delay, many early-exit strategies have been proposed where the inference is performed only through a few layers instead of the entire DNN and do not use the complete DNN~\cite{li2019edge, pacheco2021towards, wang2019adda}.} Most work in this category aims at segregating these network splits into different devices based on their computational performance~\cite{matsubara2019distilled, teerapittayanon2017distributed, goli2020migrating, zhang2021deepslicing, kang2017neurosurgeon}.  Thus, \blue{fast and localized inference using shallow portions of DL models can allow quick inference}, possibly at the cost of poorer resulting accuracy. This gives a tradeoff between result fidelity and response time. Several works have been proposed to leverage this tradeoff for multi-objective optimization, especially to reduce the frequency of SLA violations~\cite{tuli2019edgelens, yang2020task}. Other recent methods aim at exploiting the resource heterogeneity in the same network layer by splitting and placing DNNs based on user demands and host capabilities~\cite{gunasekaran2020implications}. \blue{Such methods can split DNNs and choose from different architectural choices to reach the maximum accuracy while agreeing to the latency constraints.} Other works aim at accelerating the model run-times by appropriate scheduling of a variety of DNN models on Edge-clusters~\cite{liang2020ai}. Another method, Gillis, uses a hybrid model, wherein it employs either model-compression or layer-splitting based on the application SLA demands~\cite{gillis}. The decision is taken using a reinforcement-learning model which continuously adapts in dynamic scenarios. It is a serverless-based model serving system that automatically partitions a large model across multiple serverless functions for faster inference and reduced memory footprint per function. The Gillis method employs two model partitioning algorithms that respectively achieve latency optimal serving and cost-optimal serving with SLA compliance. However, \blue{this method cannot jointly optimize both latency and SLA. Moreover, it does not consider the mobility of devices or users and hence is ineffective in efficiently managing large DNNs in mobile Edge computing environments.}  

\textbf{Splitting.} Two types of splitting strategies exist: data splitting and semantic model splitting. Data splitting splits the input data batch across multiple instances of the neural networks for parallel inference. \blue{Data splitting allows reducing the response time of inference over input data, at the cost of higher network overheads~\cite{kaplunovich2020automatic}. Semantic model splitting divides the network weights into a hierarchy of multiple groups that use a different set of features.} Here, the neural network is split based on the data semantics, producing a tree structured model that has no connection among branches of the tree, allowing parallelization of input analysis~\cite{kim2017splitnet}. Due to limited information sharing among the neural network fragments, the semantic splitting scheme gives lower accuracy in general compared to unsplit networks. Semantic splitting requires a separate training procedure where publicly available pre-trained models cannot be used. This is because a pre-trained standard neural network can be split layer-wise without affecting output semantics. For semantic splitting, we would need to first split the neural network based on data semantics and re-train the model. However, semantic splitting provides parallel task processing and hence lower inference times, more suitable for mission-critical tasks like healthcare and surveillance. Examples of such methods include ThriftyEdge~\cite{chen2018thriftyedge}, CLIO~\cite{huang2020clio}, SplitPlace~\cite{tuli2021splitplace} and SplitNet~\cite{kim2017splitnet, ushakov2018split}.

\blue{\textbf{TinyML.} This is a paradigm where the objective is to run complex deep learning models within resource constrained embedded devices~\cite{ray2021review}. Although many of the above approaches have high overlap with the methods considered in the scope of TinyML, we specifically discuss the advances in computational algorithms to augment resource management in fog environments. For instance, hyper-dimensional computing (HDC) is an approach that consumes much lower energy compared to conventional methods. Here the tensors of DNNs are mapped to higher dimensional tensors~\cite{ge2020classification}. Another approach to improve the memory footprint and minimize the read/write latencies is swapping~\cite{miao2021enabling} where DNN models are efficienctly swapped between the on-chip memory of the microcontroller and external flash memory. Another recent approach is \textit{attention condenser} that is an auxiliary neural network that learns self-attention to condense the size of the input~\cite{wong2020attendnets}. }

\subsubsection{AI Augmented Resource Provisioning}
\label{sec:deployement_provisioning}
Systematic resource provisioning is central to cost and resource-efficient computation in Fog systems. Bootstrapping resources, such as Cloud VMs or Edge nodes is time-consuming for latency-critical tasks; a key challenge is to predict future workload demands to provision resources to optimize QoS. Resource management is a key aspect of resource provisioning, which instantiates and deallocates resources based on dynamic workload demands. Most prior work aims to automate resource provisioning to optimize various performance measures such as energy consumption, cost, and task response time~\cite{tuli2021hunter,narya}. However, this problem is challenging due to the non-stationary utilization characteristics of most workloads~\cite{ebadifard2021autonomic}, requiring methods to dynamically adapt their provisioning policies. Most dynamic resource provisioning methods decouple the provisioning problem into two stages: demand prediction and decision optimization~\cite{uahs}. This is commonly referred to as the \textit{predict+optimize} framework in literature. Thus, we divide prior approaches based on their decision type.

\textbf{Demand Prediction.} Methods that forecast demands at a future state of a Fog system need data corresponding to historical workload demands on the same system. Several methods have been proposed that leverage a forecasting model. For instance, a class of methods utilize regression models such as Linear Regression (LR)~\cite{hyndman2018forecasting}, Support Vector Regression (SVR)~\cite{zhu2016pso}, Gaussian Process Regression~\cite{chen2018resource, uahs, cahs} or modular regression (Prophet)~\cite{taylor2018forecasting}. Others utilize auto-regressive models such as AutoARIMA~\cite{arima, calheiros2014workload} or other ML models like Regression Tree (RT)~\cite{xu2020hybrid}, time series decomposition (TSDec)~\cite{hyndman2018forecasting} or unobserved component model (UCM)~\cite{durbin2012time} based forecasting. Recent works utilize DNNs to perform forecasting, for instance, using LSTM neural networks~\cite{lstm}, CNNs~\cite{bega2019deepcog}, convolutional wavelet neural networks~\cite{jeddi2019water} or LSTM based GANs~\cite{yazdanian2021e2lg}. \blue{DNN based demand prediction models are known to outperform classical AI or regression based approaches~\cite{yazdanian2021e2lg,jeddi2019water}.}

\textbf{Decision Optimization.} Using a demand prediction model, several previous works optimize the provisioning decision to minimize execution costs or maximize the utilization ratio. Conventional methods often use evolutionary search strategies such as Ant Colony Optimization (ACO)~\cite{aco}, which has been shown to exhibit state-of-the-art QoS scores in recent work~\cite{cahs}. Others use Bayesian Optimization (BO)~\cite{uahs, cahs}, Particle Swarm Optimization (PSO)~\cite{zhu2016pso, chen2020self} or Genetic Algorithms~\cite{asghari2021task}. \blue{Among the different approaches, ACO and PSO are appropriate for static scenarios, whereas GA and BO are more suitable for highly dynamic settings~\cite{tuli2021cosco,asghari2021task}.}

\textbf{Hybrid Provisioning.} Other methods, such as Decision-NN, combine the prediction and optimization steps by modifying the loss function to train neural networks in conjunction with the optimization algorithm~\cite{decisionnn}. This method uses a neural network as a surrogate model to directly predict optimization objectives and uses the concept of neural network inversion, wherein the method evaluates gradients of the objective function with respect to inputs and runs optimization in the input space. \blue{However, continuous relaxation of the discrete optimization problem used in this work has been shown to adversely impact performance~\cite{uahs}.} A similar method, Semi-Direct, utilizes dynamic programming to find the optimal provisioning decision~\cite{semidirect}, but offers limited scalability with workload size. Similarly, Narya~\cite{narya} is built for mitigating VM interruptions in Cloud machines, but can be straightforwardly extended to resource provisioning. It uses a neural network as a surrogate model with a multi-armed bandit model to decide provisioning actions. \blue{However, it faces the problem of exposure bias, \textit{i.e.}, the neural model is biased to the trends in the training data and is unable to forecast in unseen cases.}

\textbf{Reactive Provisioning.} Recently, RL based methods have been proposed for reactive provisioning. For instance, Intelligent Resource Allocation Framework (iRAF)~\cite{chen2019iraf} solves the complex resource allocation problem for the collaborative mobile Edge computing (CoMEC) network using Deep Reinforcement Learning (DRL) with a multi-task objective formulation. It makes resource allocation decisions based on network states and other task characteristics such as the computing capability of devices, network quality, resource utilization, and latency requirements. iRAF automatically takes into account the network environment and makes resource allocation decisions to maximize the performance over latency and power consumption. It uses self-play training where the agent becomes its own teacher and learns over time in a self-supervised learning manner. Specifically, it uses a fully connected network (FCN) with Monte Carlo Tree Search (MCTS) to optimize the provisioning decision. Some other works, such as DDRM~\cite{chen2020deep}, focus on the integration of IoT and industrial manufacturing systems (IIoT). The authors argue that due to the limitation of computing capacity and battery, computation-intensive tasks need to be executed in the mobile Edge computing (MEC) server. %The problem is challenging in that it is dynamic and needs to maintain service continuity. It formulates the problem of joint power control and resource provisioning for MEC in IIoT. It minimizes the long-term average delay of the tasks by modeling the problem as an MDP and applying a reinforcement learning-based dynamic resource management (DDRM) algorithm to solve the formulated MDP problem. 
Another similar work~\cite{baek2020heterogeneous} focuses on optimizing the Fog nodes by selecting the suitable nodes and proper resource management while guaranteeing the QoS requirements of the users. It designs a joint task offloading and resource allocation control for heterogeneous service tasks in multi-fog nodes systems. It applies a deep recurrent Q-network (DRQN) approach to approximate the optimal value functions and applies an adjusted exploration-exploitation method to make the optimization process more efficient. Similarly, ReCARL~\cite{xu2020recarl} focuses on Cloud Radio Access Networks (CRANs). It proposes a resource allocation scheme in CRANs to improve the objective of power consumption and SLA violations of wireless users over a long time period. To do this, it uses DRL to solve a custom convex optimization problem and apply a Deep Neural Network (DNN) to approximate the action-value function. It uses two DRL agents: ReCARL-Basic (requiring limited training) and ReCARL-Hybrid (requiring deep learning training). It has been evaluated via extensive simulation to demonstrate that ReCARL achieves significant power savings in highly dynamic settings while meeting user SLA demands. Similarly, Deep Elastic Resource Provisioning (DERP)~\cite{bitsakos2018derp} uses Deep-Q learning to optimize provisioning decisions with utilization ratio as a reward for the DRL agent. Unlike Q-learning based agents that utilize a neural network to predict the expected reward for each action~\cite{sami2021ai}, recent methods also use neural networks to approximate the optimal policy. Such approaches are called policy gradient methods and include~\cite{xu2020deep, chen2021deep}. \blue{The state-of-the-art policy gradient methods outperform traditional reinforcement learning (Q learning) and Monte Carlo based approaches~\cite{xu2020deep,chen2021deep}.}

\begin{table*}[t]
    \centering
    \caption{Summary of state-of-the-art methods for AI augmented scheduling. Color scheme as per Table~\ref{tab:trends}. }
    \label{tab:scheduling}
    \resizebox{\linewidth}{!}{
    \begin{tabular}{@{}|p{1.8cm}|p{2.3cm}|c|c|c|c|>{\centering}p{1.9cm}|c|c|@{}}
\hline 
Decision Type & Category & Ref. & Method & Infra. & Benchmark & Framework / Simulator & Merits & Limitations\tabularnewline
\hline 
\hline 
\multirow{16}{2cm}{Bag-of-Tasks Scheduling} & \multirow{2}{2cm}{Maxweight Scheduling} & \cite{krishnasamy2018augmenting} & \cellcolor{pink!25}Maxweight & E+C & - & S: Custom & Fast Execution & Resp. Time\tabularnewline
\cline{3-9} \cline{4-9} \cline{5-9} \cline{6-9} \cline{7-9} \cline{8-9} \cline{9-9} 
 &  & \cite{liu2020pond} & \cellcolor{pink!25}Maxweight & E+C & Tutoring & S: Custom & Fast Execution & Resp. Time\tabularnewline
\cline{2-9} \cline{3-9} \cline{4-9} \cline{5-9} \cline{6-9} \cline{7-9} \cline{8-9} \cline{9-9} 
 & \multirow{5}{2cm}{Surrogate Modeling} & \cite{jiang2019deep} & \cellcolor{pink!25}DNN + Linear Programming & E & - & S: Custom & Energy Efficient & Accuracy\tabularnewline
\cline{3-9} \cline{4-9} \cline{5-9} \cline{6-9} \cline{7-9} \cline{8-9} \cline{9-9} 
 &  & \cite{han2018fog} & \cellcolor{magenta!25}DNN + GA & E+C & - & - & Generalizable & Overhead\tabularnewline
\cline{3-9} \cline{4-9} \cline{5-9} \cline{6-9} \cline{7-9} \cline{8-9} \cline{9-9} 
 &  & \cite{tuli2020ithermofog} & \cellcolor{magenta!25}GMM + GA & E+C & Bitbrain & S: iFogSim & Fast Execution & Overhead\tabularnewline
\cline{3-9} \cline{4-9} \cline{5-9} \cline{6-9} \cline{7-9} \cline{8-9} \cline{9-9} 
 &  & \cite{tuli2021cosco} & \cellcolor{green!25}DNN + Gradient Opt. & E+C & DeFog & F: COSCO & High QoS & Interpretability\tabularnewline
\cline{3-9} \cline{4-9} \cline{5-9} \cline{6-9} \cline{7-9} \cline{8-9} \cline{9-9} 
 &  & \cite{tuli2021hunter} & \cellcolor{green!25}GNN + Gradient Opt. & C & DeFog & F: COSCO & High QoS & Exposure Bias\tabularnewline
\cline{2-9} \cline{3-9} \cline{4-9} \cline{5-9} \cline{6-9} \cline{7-9} \cline{8-9} \cline{9-9} 
 & \multirow{4}{2cm}{Stochastic Modeling} & \cite{jamshidi2016uncertainty, bui2017energy} & \cellcolor{red!25}Gaussian Process Regression & C & Google Cluster & S: Custom & Fast Execution & Overhead\tabularnewline
\cline{3-9} \cline{4-9} \cline{5-9} \cline{6-9} \cline{7-9} \cline{8-9} \cline{9-9} 
 &  & \cite{panda2015uncertainty, jawad2018robust} & \cellcolor{yellow!25}Robust Search & C & Custom & - & Energy Efficient & Overhead\tabularnewline
\cline{3-9} \cline{4-9} \cline{5-9} \cline{6-9} \cline{7-9} \cline{8-9} \cline{9-9} 
 &  & \cite{alelaiwi2019efficient} & \cellcolor{yellow!25}DBN + Search & E+C & - & S: Custom & Fast Execution & Accuracy\tabularnewline
\cline{3-9} \cline{4-9} \cline{5-9} \cline{6-9} \cline{7-9} \cline{8-9} \cline{9-9} 
 &  & \cite{tuli2021gosh} & \cellcolor{green!25}NPN + Gradient Opt. & E+C & DeFog & F: COSCO & High QoS & Interpretability\tabularnewline
\cline{2-9} \cline{3-9} \cline{4-9} \cline{5-9} \cline{6-9} \cline{7-9} \cline{8-9} \cline{9-9} 
 & \multirow{4}{2cm}{Reinforcement Learning} & \cite{tang2018migration, li2019deep} & \cellcolor{blue!25}Deep Q Learning & E+C & Crawdad & S: Cloudsim & Generalizable & Scalability\tabularnewline
\cline{3-9} \cline{4-9} \cline{5-9} \cline{6-9} \cline{7-9} \cline{8-9} \cline{9-9} 
 &  & \cite{wang2021energy} & \cellcolor{cyan!25}Minimax Q Learning & C & Traffic & S: Clousim & Generalizable & Scalability\tabularnewline
\cline{3-9} \cline{4-9} \cline{5-9} \cline{6-9} \cline{7-9} \cline{8-9} \cline{9-9} 
 &  & \cite{sheng2021deep} & \cellcolor{Fuchsia!25}Policy Gradient Learning & E & - & S: Custom & High QoS & Adaptability\tabularnewline
\cline{3-9} \cline{4-9} \cline{5-9} \cline{6-9} \cline{7-9} \cline{8-9} \cline{9-9} 
 &  & \cite{tuli2020dynamic, cheng2021multi} & \cellcolor{Fuchsia!25}Policy Gradient Learning & E & Bitbrain & S: iFogSim & High QoS & Adaptability\tabularnewline
\cline{2-9} \cline{3-9} \cline{4-9} \cline{5-9} \cline{6-9} \cline{7-9} \cline{8-9} \cline{9-9} 
 & \multirow{1}{2.3cm}{Co-Simulation} & \cite{tuli2021cosco, tuli2021gosh} & \cellcolor{LimeGreen!25}FCN + Co-Simulation & E+C & DeFog & F: COSCO & Generalizable & Overhead\tabularnewline
\hline 
\multirow{14}{2cm}{Workflow Scheduling} & \multirow{3}{2cm}{Meta-Heuristic Methods} & \cite{impso} & \cellcolor{magenta!25}Particle Swarm Optimization & C & WfCommons & F: AWS IoT & Generalizable & Low QoS\tabularnewline
\cline{3-9} \cline{4-9} \cline{5-9} \cline{6-9} \cline{7-9} \cline{8-9} \cline{9-9} 
 &  & \cite{huang2020ant} & \cellcolor{magenta!25}Ant Colony Optimization & E+C & WfCommons & F: AWS IoT & Generalizable & Low QoS\tabularnewline
\cline{3-9} \cline{4-9} \cline{5-9} \cline{6-9} \cline{7-9} \cline{8-9} \cline{9-9} 
 &  & \cite{ghanavati2020energy} & \cellcolor{magenta!25}Ant Mating Optimization & E+C & WfCommons & S: Custom & Energy Efficient & Overhead\tabularnewline
\cline{2-9} \cline{3-9} \cline{4-9} \cline{5-9} \cline{6-9} \cline{7-9} \cline{8-9} \cline{9-9} 
 & \multirow{3}{2cm}{Surrogate Optimization} & \cite{dnsga} & \cellcolor{magenta!25}DNN + GA & C & WfCommons & - & Generalizable & Overhead\tabularnewline
\cline{3-9} \cline{4-9} \cline{5-9} \cline{6-9} \cline{7-9} \cline{8-9} \cline{9-9} 
 &  & \cite{esvr} & \cellcolor{magenta!25}DNN + GA & C & WfCommons & S: Cloudsim & Cost Efficient & Overhead\tabularnewline
\cline{3-9} \cline{4-9} \cline{5-9} \cline{6-9} \cline{7-9} \cline{8-9} \cline{9-9} 
 &  & \cite{tuli2021mcds} & \cellcolor{green!25}DNN + Gradient Opt. & E+C & WfCommons & F: COSCO & High QoS & Interpretability\tabularnewline
\cline{2-9} \cline{3-9} \cline{4-9} \cline{5-9} \cline{6-9} \cline{7-9} \cline{8-9} \cline{9-9} 
 & \multirow{1}{2cm}{Game Theory} & \cite{closure} & \cellcolor{pink!25}Attack-Defense Model & C & WfCommons & F: Openstack & Cost Efficient & Overhead\tabularnewline
\cline{2-9} \cline{3-9} \cline{4-9} \cline{5-9} \cline{6-9} \cline{7-9} \cline{8-9} \cline{9-9} 
 & \multirow{4}{2cm}{Reinforcement Learning} & \cite{wang2019multi} & \cellcolor{blue!25}Deep Q Learning & C & WfCommons & F: AWS IoT & Generalizable & Scalability\tabularnewline
\cline{3-9} \cline{4-9} \cline{5-9} \cline{6-9} \cline{7-9} \cline{8-9} \cline{9-9} 
 &  & \cite{kaur2020deep} & \cellcolor{blue!25}Deep Q Learning & C & WfCommons & S: Clousim{*} & Generalizable & Scalability\tabularnewline
\cline{3-9} \cline{4-9} \cline{5-9} \cline{6-9} \cline{7-9} \cline{8-9} \cline{9-9} 
 &  & \cite{ghosal2020deep} & \cellcolor{Fuchsia!25}Policy Gradient Learning & E+C & - & - & High QoS & Adaptability\tabularnewline
\cline{3-9} \cline{4-9} \cline{5-9} \cline{6-9} \cline{7-9} \cline{8-9} \cline{9-9} 
 &  & \cite{hu2019learning} & \cellcolor{Fuchsia!25}Policy Gradient Learning & C & WfCommons & F: Custom & High QoS & Adaptability\tabularnewline
\cline{2-9} \cline{3-9} \cline{4-9} \cline{5-9} \cline{6-9} \cline{7-9} \cline{8-9} \cline{9-9} 
 & \multirow{3}{2cm}{Other} & \cite{feng2019content} & \cellcolor{OliveGreen!25}LSTM & E+C & Custom & - & Generalizable & Scalability\tabularnewline
\cline{3-9} \cline{4-9} \cline{5-9} \cline{6-9} \cline{7-9} \cline{8-9} \cline{9-9} 
 &  & \cite{wang2020intelligent} & \cellcolor{red!25}Support Vector Regression & E & - & S: Custom & Fast Execution & Scalability\tabularnewline
\cline{3-9} \cline{4-9} \cline{5-9} \cline{6-9} \cline{7-9} \cline{8-9} \cline{9-9} 
 &  & \cite{alsurdeh2021hybrid} & \cellcolor{yellow!25}Gradient Descent Search & E+C & Custom & S: CloudSim & High QoS & Overhead\tabularnewline
\hline 
\end{tabular}}
\end{table*}

\subsection{AI Augmented Scheduling}
\label{sec:scheduling}

A summary of recent AI-augmented scheduling methods is presented in Table~\ref{tab:scheduling}.

\subsubsection{AI Augmented Scheduling of Bag-Of-Tasks}
\label{sec:scheduling_bot}

QoS-aware placement of IoT applications requires reaching a tradeoff among multiple conflicting QoS parameters such as response time, cost and energy. In the \textit{bag-of-task} workload model, each task can be independently scheduled.

\textbf{MaxWeight-Scheduling.} Over the years, many scheduling approaches have turned to utilize MaxWeight based techniques due to its theoretical guarantees and the ability to reduce the frequency of resource contention~\cite{liu2020pond, krishnasamy2018augmenting}. For instance, the pessimistic-optimistic online dispatch approach, POND, is a variant of the MaxWeight approach~\cite{liu2020pond}. POND formulates the scheduling problem as a constrained optimization objective with unknown dispatch, arrival and reward distributions, such that each Fog node has a virtual queue to track violation counts. It uses an Upper-Confidence Bound (UCB) based exploration strategy~\cite{auer2002finite} with the final decisions being made with the MaxWeight weights as the expected reward value of each scheduling decision. \blue{However, prior work has demonstrated that MaxWeight policies suffer from instability in dynamic workloads, high delays and inefficiency in modeling large-scale Fog networks~\cite{bae2019beyond, van2009instability, van2013inefficiency}. MaxWeight schedulers are also known to have high wait times due to their inability to adapt to volatile workload settings~\cite{tuli2021cosco}.}

\textbf{Surrogate Modeling.} Most classical research in this area employs meta-heuristic algorithms with a DNN or regression model as a surrogate that approximates QoS of a given system state. This is due to their generic formulation and ease of implementation. For instance, prior works have shown that evolutionary-based methods, and generally gradient-free approaches, perform well in dynamic scenarios~\cite{wang2019empowering, han2018fog, wang2020effective, tuli2020ithermofog}. Some works use a combination of a DNN surrogate, and classical optimization techniques such as mixed-integer linear programming (MILP)~\cite{jiang2019deep}. Evolutionary approaches such as genetic algorithms (GA) lie in the domain of gradient-free optimization methods. The GA method schedules workloads using a neural model to approximate the objective value and a genetic algorithm to reach the optimal decision~\cite{han2018fog}. Such methods use either analytical models\cite{wang2019empowering}, Gaussian Mixture Model (GMM) or polynomial approximators~\cite{tuli2020ithermofog} or neural networks~\cite{han2018fog} to predict system QoS for a given scheduling decision and input Fog state. Typically, such approaches run a search scheme with non-local jumps, due to cross-over and mutation-like operations, to converge towards an optimum. \blue{However, gradient-free methods are known to take much longer to converge~\cite{bogolubsky2016learning} and are not as scalable~\cite{rios2013derivative} as gradient-based methods. Moreover, non-local jumps can significantly change the scheduling decision, leading to a high number of preemptive task migrations.} This entails checkpointing the running task, migrating it to another Fog node and resuming its execution on the new node~\cite{engelmann2009proactive}. This can give rise to high migration overheads, subsequently increasing the average task response times and SLA violation rates. Furthermore, prior work also establishes that neural approximators can precisely model the gradients of the objective function with respect to input using back-propagation~\cite{nguyen1999approximation}. Now, although such works use these gradients with respect to input for solving differential equations, they can also be applied for gradient-based optimizations. \blue{However, even with the advantages of scalability and quick convergence to optima, few prior works use gradient-based methods as neural approximators are not consistent with the convexity/concavity requirements of such methods~\cite{nandi2001artificial}. This problem is alleviated by momentum and annealing in schedulers like GOBI and GOSH~\cite{tuli2021cosco, tuli2021gosh}.} Such methods take the scheduling decision and state of the Fog system as resource utilization characteristics of workloads and Fog nodes and output a QoS estimate. Using backpropagation to input, \textit{i.e.}, fixing the neural network parameters and updating the scheduling decision based on the gradient of DNN output, these methods find the optimal scheduling decisions. Other schedulers, like HUNTER~\cite{tuli2021hunter}, model the input scheduling decision as a graph and use Graph Neural Networks (GNNs), facilitating the inference by capturing the correlations across workloads and Fog nodes. However, with such models, as they run black-box optimization steps, the interpretability of their outputs is low. Further, continuous approximation of a discrete optimization problem is known to give sub-optimal decisions in some cases~\cite{miranda2018constraint}.

\textbf{Stochastic Modeling:} Another type of models that approximate system QoS is stochastic surrogate models. These include Heteroscedastic Gaussian Processes to approximate the distribution of the QoS metrics instead of giving only a static output for a given input state~\cite{jamshidi2016uncertainty, bui2017energy, panda2015uncertainty}. Similarly, prior works also predict the mean and variance estimates of system QoS based on historical data to perform robust and safe decision optimization~\cite{panda2015uncertainty, jawad2018robust} or use error-based exploration~\cite{jamshidi2016uncertainty}. Other methods use Deep Belief Networks (DBN) for response-time predictions, which are used to make prompt offloading decisions under mobility and fluctuating resource demands~\cite{alelaiwi2019efficient}. \blue{Typically, due to the poor modeling accuracy of Gaussian Processes, they cannot perform well in complex environments like heterogeneous Fog environments. Hence, more sophisticated models like Bayesian Neural Networks (BNNs) to additionally model the stochasticity in the QoS metrics~\cite{jawad2018robust, wu2020accelerating}.} Recent state-of-the-art methods also rely on Natural Parameter Networks (NPNs) that allow using arbitrary exponential family of distributions to model the weights and parameters of a neural networks~\cite{tuli2021gosh}.

\textbf{Reinforcement Learning models:} Recently, reinforcement learning based methods have shown themselves to be robust and versatile to diverse workload characteristics and complex Fog setups~\cite{tuli2020dynamic, tang2018migration, basu2019learn, gazori2019saving}. Such methods use a Markovian assumption of state which is the scheduling decision at each interval. Based on new observations of reward signals, they explore or exploit their knowledge of the state-space to converge to an optimal decision. Recent methods, such as DQLCM~\cite{tang2018migration} and DeepRM~\cite{li2019deep}, model the container migration problem as a multi-dimensional Markov Decision Process (MDP) and use a deep-reinforcement learning strategy, namely deep Q-Learning to schedule workloads in a heterogeneous Fog environment. Another similar method, SDAEM-MMQ~\cite{wang2021energy} uses a stacked denoising autoencoder with minimax Q learning for accurate Q estimates and robust optimization. Policy gradient methods, such as~\cite{sheng2021deep}, train a DNN to directly predict the optimal scheduling decision instead of Q values. A recent method, Asynchronous Advantage Actor-Critic (A3C), is a policy gradient method that schedules workloads using an actor-critic pair of DNN agents~\cite{tuli2020dynamic}. This approach uses Residual Recurrent Neural Networks (R2N2) to predict the expected reward for each action \emph{i.e.}, scheduling decision and tries to optimize the cumulative reward signal. Another similar method, Multi-Agent Deep Deterministic Policy Gradient (MADDPG)~\cite{cheng2021multi} formulates the decision optimization problem as a stochastic game among multiple RL agents to reach to an optimal schedule. However, such methods are still slow to adapt to real-world application scenarios~\cite{tuli2021cosco}. This leads to higher wait times and subsequently high response times and SLA violations, leading to poor scalability with workload or the number of nodes in the Fog system.

\textbf{Coupled Optimization.} Finally, coupled or symbiotic simulation and model-based control have long been used in the modeling and optimization of distributed systems~\cite{onggo2021combining, bosmans2019testing, onggo2018symbiotic}. Many prior works have used hybrid simulation models to optimize decision-making in dynamic systems~\cite{mustafee2015hybrid, onggo2018symbiotic}. To achieve this, they monitor, analyze, plan and execute decisions using previous knowledge-base corpora (MAPE-k)~\cite{gill2019transformative}. However, such works use this to facilitate search methods and not generate additional data to aid the decision-making of an AI model. Recent methods, such as GOBI*~\cite{tuli2021cosco}, use an interleaved decision optimization and co-simulation to run an interactive dynamic between the different levels of fidelity, \textit{i.e.}, simulation and surrogate, to optimize QoS. %Although GOBI* provides distinct advantages compared to the state-of-the-art, it also faces drawbacks. First, the model requires a neural approximator to be pre-trained on several hours of workload traces generated using a random allocation-based scheduler. This might be infeasible for application scenarios where swift deployment is required. Second, the deployment environment might be very different from the setup used to generate the training data for the model. Even though the model is able to adapt to completely different setups, experiments show that GOBI* lacks the agility required for QoS efficient scheduling. However, even with techniques like momentum, our experiments show that gradient-based optimization can take a large number of iterations to converge due to many saddle points in the QoS hyper-surface. Considering these shortcomings, GOBI* has been augmented with higher-order optimization and stochastic modeling to present the GOSH* scheduler~\cite{tuli2021gosh}.
A similar work~\cite{onggo2018symbiotic}, presents the notion of \textit{symbiotic simulation} that aims to feed in the resource characteristics related data to a co-simulated model for optimizing resource management related decisions using ML techniques. Another similar work, EDSS~\cite{onggo2021combining}, uses a co-simulator to estimate the effects of various resource scheduling decisions from an ML model and choose the one with highest QoS. \blue{Running a co-simulator gives another estimate of system QoS, solving two problems at once: the problem of exposure bias to training data as well as the data saturation problem. The former arises due to the surrogate models being trained on a set of pre-collected execution traces, wherein the system characteristics might be different from those at test time. The latter arises due to the limited diversity in training data such that even with more datapoints, the performance of the DNN does not improve.}

\subsubsection{AI Augmented Scheduling of Workflows}
\label{sec:scheduling_workflows}

Workflow like applications typically have precedence constraints of the form of a DAG that must be adhered to when scheduling such applications. These workloads could be of the form of a layer or semantic split neural models (see Section~\ref{sec:deployment_dnn}) or other scientific workflow applications~\cite{gill2019transformative}.

\textbf{Meta-Heuristic methods.} This class of methods leverages high-level problem independent algorithms to find the optimal scheduling decision for the workflows. Most state-of-the-art approaches belong to this category. Among these, many use variants of the PSO optimization technique~\cite{impso}. One such technique is the immune-based particle swarm optimization (IMPSO) method~\cite{impso}. It uses candidate affinity to prevent poor candidates from being discarded in subsequent iterations, allowing it to surpass other PSO-based methods in terms of execution costs and average response time. Other techniques, categorized commonly as list scheduling, use metrics like earliest finish time, critical path, and dynamic resource utilization levels~\cite{adhikari2019survey}. However, list scheduling performs poorly in settings with non-preemptable jobs and heterogeneous requirements or machines~\cite{adhikari2019survey}. Others include ACO, such as~\cite{huang2020ant}. Such a technique starts with several random or heuristically initialized candidate solutions. Each candidate is iteratively optimized, moving it slightly in the state-space where the optimization objective tends to increase. \blue{Such methods aim to reach a balance between makespan-service spread, makespan-energy and makespan-cost, respectively.} Further, novel bio-inspired meta-heuristic algorithms are also introduced to solve Fog application scheduling simultaneously considering multiple objectives. For instance, the Ant Mating Optimization (AMO)~\cite{ghanavati2020energy}, aims to minimize the total system makespan and energy consumption for Fog task scheduling. 

\textbf{Surrogate Optimization.} Other recent methods use genetic algorithms to optimize the scheduling decision, again using a DNN as a surrogate model~\cite{dnsga, esvr}. Again, due to non-local jumps in the search space, such methods typically lead to better QoS estimates, at the cost of higher task migration overheads~\cite{tuli2021cosco}. Recent techniques, such as ESVR~\cite{esvr}, initialize its candidate population using the Heterogeneous Earliest Finish Time (HEFT) heuristic and optimize using the crossover-mutation scheme. To account for volatility in the system, ESVR continuously fine-tunes the neural network surrogate using the latest workload traces and host characteristics~\cite{esvr}. A similar technique is DNSGA~\cite{dnsga} that uses a multi-objective optimization method that uses a Pareto Optimal Front (POF) aware approach that prevents the set of candidates from converging to the same optima~\cite{dnsga}. \blue{Prior work shows that these two methods out-perform previously proposed genetic algorithms-based techniques~\cite{dnsga, esvr}.} However, in the case of long-running workflows, having a short-term QoS estimate is detrimental to the system performance as it leads to myopic optimization. To tackle this, recent methods, such as Monte-Carlo Deep Surrogate (MCDS)~\cite{tuli2021mcds}, trains a DNN to generate long-term QoS estimates by running multiple Monte-Carlo simulations on a co-simulator. This not only helps in long-term optimization, but also facilitates stable learning.

\textbf{Game-Theoretic Modeling.} Another recently proposed workflow scheduling model, namely Closure, uses an attack-defense game-theoretic formulation~\cite{closure}. Unlike other schemes that assume mostly homogeneous resources, \blue{Closure has been shown to efficiently manage heterogeneous devices by calculating the Nash Equilibrium of the attack-defense game model}. This is crucial in Edge-cloud environments where there are contrasting resource capacities of Edge and Cloud.

\textbf{Reinforcement Learning.} Some methods constrain the action space of the MDP formulation to exclude scheduling decisions that violate the precedence constraints set by the incoming workloads. These include removing infeasible actions from the action set at each state of the MDP~\cite{wang2019multi} for Deep Q Networks (DQN) or masking the policy likelihood scores in policy gradient methods~\cite{hu2019learning, ghosal2020deep}. For instance, DQ-HEFT~\cite{kaur2020deep} superimposes the task order over the reward function to ensure that the Q-learning model converges to an optimal scheduling decision.

\textbf{Other.} Many prior works utilize other augmentation strategies in tandem with AI. For instance, some works optimize the Fog network. Examples include~\cite{jalali2019dynamic, jalali2017cognitive}, which introduce cognitive Edge gateways that use machine learning (regression and ensemble models) to automatically learn the best allocation for each task based on the Fog environment status and performance requirements of the tasks. Similarly, other methods~\cite{wang2020intelligent} propose an intelligent task offloading algorithm to synergistically run them on Edge and Cloud platforms.\blue{ Their dynamic switching algorithm groups applications using a support vector machine based approach to improve the performance in terms of delay and energy consumption.} Other methods such as gradient descent search~\cite{alsurdeh2021hybrid} has been adopted for hybrid workflow scheduling in Edge and cloud computing to optimize execution time and monetary cost.  %Some prior works~\cite{jiang2020deep,feng2019content,yu2021privacy} use AI-based prediction mechanisms to generate optimal caching strategies. \cite{feng2019content} trains a simplified bidirectional long short-term memory (Bi-LSTM) network to make accurate and adaptive popularity predictions to improve the cache hit rate. \cite{jiang2020deep} propose a proactive-reactive caching policy that uses deep learning and k-Nearest Neighbor (kNN) to classify popularity trends and track user dynamics to make caching decisions. \cite{yu2021privacy} address the caching problem while considering data security violations caused by centralized training of the models. A Federated learning-based caching scheme is proposed where IoT devices use local data and train a shared model to predict content popularity.

\begin{table*}[t]
    \centering
    \caption{Summary of state-of-the-art methods for AI augmented maintenance. Color scheme as per Table~\ref{tab:trends}.}
    \label{tab:maintenance}
    \resizebox{\linewidth}{!}{
\begin{tabular}{|>{\raggedright}p{1.7cm}|>{\raggedright}m{2.cm}|c|c|c|c|>{\centering}p{2cm}|c|c|}
\hline 
Decision Type & Category & Ref. & Method & Infra. & Benchmark & Framework / Simulator & Merits & Limitations\tabularnewline
\hline 
\hline 
\multirow{13}{1.3cm}{Fault Detection and Prediction} & \multirow{4}{2cm}{Unsupervised Reconstruction Models} & \cite{lstm_ndt, omnianomaly, lstm_vae} & \cellcolor{OliveGreen!25}LSTM & E+C & SMD & S: Custom & Generalizable & Overhead\tabularnewline
\cline{3-9} \cline{4-9} \cline{5-9} \cline{6-9} \cline{7-9} \cline{8-9} \cline{9-9} 
 &  & \cite{mscred, cae_m} & \cellcolor{SeaGreen!25}ConvLSTM & E+C & Custom & S: Custom & High Accuracy & Scalability\tabularnewline
\cline{3-9} \cline{4-9} \cline{5-9} \cline{6-9} \cline{7-9} \cline{8-9} \cline{9-9} 
 &  & \cite{dagmm} & \cellcolor{red!25}DNN + GMM & E+C & Custom & S: Custom & High Accuracy & Overhead\tabularnewline
\cline{3-9} \cline{4-9} \cline{5-9} \cline{6-9} \cline{7-9} \cline{8-9} \cline{9-9} 
 &  & \cite{mtad_gat, gdn} & \cellcolor{gray!25}Graph Neural Network & E+C & SWaT & S: Custom & High Accuracy & Overhead\tabularnewline
\cline{2-9} \cline{3-9} \cline{4-9} \cline{5-9} \cline{6-9} \cline{7-9} \cline{8-9} \cline{9-9} 
 & \multirow{6}{2cm}{Generative Models} & \cite{mad_gan, stepgan} & \cellcolor{gray!25}Generative Adversarial Nets & E+C & SWaT & - & Fast and Scalable & Exposure Bias\tabularnewline
\cline{3-9} \cline{4-9} \cline{5-9} \cline{6-9} \cline{7-9} \cline{8-9} \cline{9-9} 
 &  & \cite{audibert2020usad, gan2020sage} & \cellcolor{LimeGreen!25}VAE & C & - & F: Custom & Memory Efficient & Scalability\tabularnewline
\cline{3-9} \cline{4-9} \cline{5-9} \cline{6-9} \cline{7-9} \cline{8-9} \cline{9-9} 
 &  & \cite{topomad, girish2021anomaly} & \cellcolor{OliveGreen!25}GNN + LSTM + VAE & C & HDFS & F: Custom & High Accuracy & Execution Time\tabularnewline
\cline{3-9} \cline{4-9} \cline{5-9} \cline{6-9} \cline{7-9} \cline{8-9} \cline{9-9} 
 &  & \cite{chouliaras2021detecting} & \cellcolor{OliveGreen!25}LSTM + VAE & C & - & - & Generalizable & Overhead\tabularnewline
\cline{3-9} \cline{4-9} \cline{5-9} \cline{6-9} \cline{7-9} \cline{8-9} \cline{9-9} 
 &  & \cite{huang2020hitanomaly} & \cellcolor{gray!25}Transformers & E+C & HDFS & F: Openstack & Fast Execution & Generalizability\tabularnewline
\cline{3-9} \cline{4-9} \cline{5-9} \cline{6-9} \cline{7-9} \cline{8-9} \cline{9-9} 
 &  & \cite{tuli2021generative} & \cellcolor{orange!25}Neural Design & E & SMD & F: COSCO & Memory Efficient & Overhead\tabularnewline
\cline{2-9} \cline{3-9} \cline{4-9} \cline{5-9} \cline{6-9} \cline{7-9} \cline{8-9} \cline{9-9} 
 & \multirow{3}{2cm}{Clustering Methods} & \cite{won2021performance} & \cellcolor{gray!25}Few Shot Learning & C & - & F: Openstack & Fast Execution & Accuracy\tabularnewline
\cline{3-9} \cline{4-9} \cline{5-9} \cline{6-9} \cline{7-9} \cline{8-9} \cline{9-9} 
 &  & \cite{li2018comparison} & \cellcolor{gray!25}Fuzzy Clustering & E+C & - & - & Fast Execution & Accuracy\tabularnewline
\cline{3-9} \cline{4-9} \cline{5-9} \cline{6-9} \cline{7-9} \cline{8-9} \cline{9-9} 
 &  & \cite{awgg} & \cellcolor{gray!25}VAE + Fuzzy Clustering & E+C & Custom & S: Custom & Fast Execution & Accuracy\tabularnewline
\hline 
\multirow{19}{2cm}{Fault Remediation} & \multirow{6}{2cm}{Fault-Tolerant Scheduling} & \cite{lbos} & \cellcolor{cyan!25}Q Learning & E+C & MHealth & - & Generalizable & Scalability\tabularnewline
\cline{3-9} \cline{4-9} \cline{5-9} \cline{6-9} \cline{7-9} \cline{8-9} \cline{9-9} 
 &  & \cite{elbs} & \cellcolor{green!25}Deep Surrogate Optimization & E+C & MHealth & S: iFogSim & Cost Efficient & Interpretability\tabularnewline
\cline{3-9} \cline{4-9} \cline{5-9} \cline{6-9} \cline{7-9} \cline{8-9} \cline{9-9} 
 &  & \cite{pcft} & \cellcolor{magenta!25}Particle Swarm Optimization & C & - & S: Cloudsim{*} & Memory Efficient & Scalability\tabularnewline
\cline{3-9} \cline{4-9} \cline{5-9} \cline{6-9} \cline{7-9} \cline{8-9} \cline{9-9} 
 &  & \cite{satpathy2018crow} & \cellcolor{yellow!25}Crow Search & C & Custom & S: Cloudsim & Energy Efficient & Scalability\tabularnewline
\cline{3-9} \cline{4-9} \cline{5-9} \cline{6-9} \cline{7-9} \cline{8-9} \cline{9-9} 
 &  & \cite{wang2021ddqp} & \cellcolor{blue!25}Deep Q Learning & E+C & - & S: Custom & High Throughput & Scalability\tabularnewline
\cline{3-9} \cline{4-9} \cline{5-9} \cline{6-9} \cline{7-9} \cline{8-9} \cline{9-9} 
 &  & \cite{tuli2021pregan} & \cellcolor{gray!25}Generative Adversarial Nets & E & DeFog & F: COSCO & High QoS & Data Hungry\tabularnewline
\cline{2-9} \cline{3-9} \cline{4-9} \cline{5-9} \cline{6-9} \cline{7-9} \cline{8-9} \cline{9-9} 
 & \multirow{4}{2cm}{Load Balancing} & \cite{jan2021ai} & \cellcolor{magenta!25}Particle Swarm Optimization & E+C & - & S: iFogSim & Fast Execution & Scalability\tabularnewline
\cline{3-9} \cline{4-9} \cline{5-9} \cline{6-9} \cline{7-9} \cline{8-9} \cline{9-9} 
 &  & \cite{kaur2021focalb} & \cellcolor{magenta!25}Grey Wolf Optimization & E+C & WfCommons & S: iFogSim{*} & Cost Efficient & Scalability\tabularnewline
\cline{3-9} \cline{4-9} \cline{5-9} \cline{6-9} \cline{7-9} \cline{8-9} \cline{9-9} 
 &  & \cite{marahatta2020pefs} & \cellcolor{yellow!25}DNN + Search & C & Google Cluster & S: Cloudsim & High Util. Ratio & Interpretability\tabularnewline
\cline{3-9} \cline{4-9} \cline{5-9} \cline{6-9} \cline{7-9} \cline{8-9} \cline{9-9} 
 &  & \cite{arabnejad2017fuzzy} & \cellcolor{gray!25}Fuzzy Logic + Search & C & - & F: Openstack & Low Overhead & Low QoS\tabularnewline
\cline{2-9} \cline{3-9} \cline{4-9} \cline{5-9} \cline{6-9} \cline{7-9} \cline{8-9} \cline{9-9} 
 & \multirow{7}{2cm}{Scaling} & \cite{etemadi2021cost} & \cellcolor{OliveGreen!25}RNN & E+C & Custom & S: iFogSim & Low Overhead & Scalability\tabularnewline
\cline{3-9} \cline{4-9} \cline{5-9} \cline{6-9} \cline{7-9} \cline{8-9} \cline{9-9} 
 &  & \cite{abdullah2020predictive} & \cellcolor{red!25}Decision Tree Regression & E+C & Custom & F: Custom & Fast Execution & Scalability\tabularnewline
\cline{3-9} \cline{4-9} \cline{5-9} \cline{6-9} \cline{7-9} \cline{8-9} \cline{9-9} 
 &  & \cite{etemadi2020learning} & \cellcolor{gray!25}NAR Network & E+C & T-Drive & S: iFogSim{*} & Cost Efficient & Scalability\tabularnewline
\cline{3-9} \cline{4-9} \cline{5-9} \cline{6-9} \cline{7-9} \cline{8-9} \cline{9-9} 
 &  & \cite{li2020heterogeneity} & \cellcolor{red!25}DNN + ARIMA & E+C & Custom & - & Cost Efficient & Scalability\tabularnewline
\cline{3-9} \cline{4-9} \cline{5-9} \cline{6-9} \cline{7-9} \cline{8-9} \cline{9-9} 
 &  & \cite{DBLP:journals/corr/abs-2103-06381} & \cellcolor{gray!25}Fuzzy Logic + Search & E+C & - & S: Cloudsim{*} & Fast and Scalable & Scalability\tabularnewline
\cline{3-9} \cline{4-9} \cline{5-9} \cline{6-9} \cline{7-9} \cline{8-9} \cline{9-9} 
 &  & \cite{9303420} & \cellcolor{gray!25}Dynamic Bayesian Network & E & SETI & - & Fast Execution & Interpretability\tabularnewline
\cline{3-9} \cline{4-9} \cline{5-9} \cline{6-9} \cline{7-9} \cline{8-9} \cline{9-9} 
 &  & \cite{aral2017quality} & \cellcolor{gray!25}Bayesian Neural Network & E & Custom & - & Fast Execution & Interpretability\tabularnewline
\cline{2-9} \cline{3-9} \cline{4-9} \cline{5-9} \cline{6-9} \cline{7-9} \cline{8-9} \cline{9-9} 
 & \multirow{2}{2cm}{Straggler Analysis} & \cite{8567669} & \cellcolor{gray!25}Dynamic Bayesian Network & E & SETI & - & Fast Execution & Interpretability\tabularnewline
\cline{3-9} \cline{4-9} \cline{5-9} \cline{6-9} \cline{7-9} \cline{8-9} \cline{9-9} 
 &  & \cite{tuli2021start} & \cellcolor{OliveGreen!25}VAE + LSTM & C & PlanetLab & S: Cloudsim & High QoS & Overhead\tabularnewline
\hline 
\end{tabular}}
\end{table*}

\subsection{AI Augmented Maintenance}

In this work, we focus on the aspect of maintaining Fog systems using resource management techniques, particularly concerned with fault tolerance, resilience and remediation. Resilience is crucial when utilizing AI for resource management, as corrupted computation from failed nodes can lead to ML systems having erroneous behavior. Such errors can be fatal in some scenarios, such as autonomous driving and medical predictions. We measure system resilience with three metrics. 
\begin{enumerate}[leftmargin=*]
    \item \textit{Resource Contention:} Stressful workloads tend to overwhelm the resource capacities of the Edge or Cloud nodes, leading to competition among workloads for resources. This competition can cause failures due to inefficient resource scheduling, resulting in outages. The most common way of mitigating this is by proactive resource provisioning that ensures sufficient resources are available for incoming workloads before they arrive. However, it is crucial to eschew the over-provisioning of resources to avoid system under-utilization or resource wastage in Fog systems.
    \item \textit{Service Availability:} It is possible that the node performing a crucial computation fails due to hardware or software faults. This disrupts the service provided to the user and is a critical metric to measure system reliability. It is usually addressed by having multiple Fog nodes involved in the processing of the same application so that one can take over if the other fails. This resilience concept is known as hot-standby, where the backup resources are called fallback nodes~\cite{zhao2020distributed}. However, this metric trades off with energy and cost as application replication leads to redundant computations, leading to inefficiency.
    \item \textit{Security and Privacy:} Fog systems must also ensure data resilience to ensure that data is not compromised by malicious attacks~\cite{zhang2019serious}. This includes data integrity, \textit{i.e.}, resilience to data corruption, and data confidentiality, \textit{i.e.}, sensitive data remains hidden from malicious entities. To avoid data corruption and stealing, technologies such as encryption~\cite{bonawitz2017practical}, differential privacy~\cite{abadi2016deep}, detection~\cite{preuveneers2018chained} are used.
\end{enumerate} 
A summary of resilience methods for Fog systems in presented in Table~\ref{tab:maintenance}.

\subsubsection{AI Augmented Fault-Detection and Prediction}
\label{sec:maintenance_pred_detect}
Several machine learning algorithms have been proposed for fault detection and prediction in Edge Cloud environments. They have proposed a framework that includes time-series data collection and data pre-processing components for the training of DNNs.  

\textbf{Unsupervised Reconstruction Models.} Majority of prior work proposes reconstruction-based methods that predominantly aim to encapsulate the temporal trends and predict the time-series system data in an unsupervised fashion, then use the deviation of the prediction with the ground-truth data as anomaly scores. In such methods, the time-series system data may correspond to utilization characteristics of the running workloads in a Fog system. One such method, LSTM-NDT~\cite{lstm_ndt}, relies on an LSTM to forecast data for the next timestamp. This work also proposes a non-parametric dynamic error thresholding (NDT) strategy to set a threshold for anomaly labeling using moving averages of the error sequence. A similar work, Omnianomaly~\cite{omnianomaly}, uses a stochastic recurrent neural network (similar to an LSTM-Variational Autoencoder~\cite{lstm_vae}) and a planar normalizing flow to generate reconstruction probabilities. It also proposes an adjusted Peak-Over-Threshold (POT) method for automated anomaly threshold selection that outperforms the previously used NDT approach. This work led to a significant performance leap compared to the prior art, but at the expense of high training times. The Multi-Scale Convectional Recursive Encoder Decoder (MSCRED)~\cite{mscred} converts an input sequence window into a normalized two-dimensional image and then passes it through a ConvLSTM layer. This method is able to capture more complex inter-modal correlations and temporal information, however is unable to generalize to settings with insufficient training data. The CAE-M~\cite{cae_m} uses a convolutional autoencoding memory network, similar to MSCRED. It passes the time-series through a CNN with the output being processed by bidirectional LSTMs to capture long-term temporal trends. Such recurrent neural network-based models have been shown to have high computation costs and low scalability for high dimensional datasets~\cite{audibert2020usad}. The DAGMM~\cite{dagmm} method uses a deep autoencoding Gaussian mixture model for dimension reduction in the feature space and recurrent networks for temporal modeling. This work predicts an output using a mixture of Gaussians, where the parameters of each Gaussian are given by a deep neural model. \blue{However, it still is slow and unable to explicitly utilize inter-modal correlations~\cite{gdn}.} The Graph Deviation Network (GDN) approach learns a graph of relationships between data modes and uses attention-based forecasting and deviation scoring to output anomaly scores. MTAD-GAT~\cite{mtad_gat} uses a graph-attention network to model both feature and temporal correlations and pass it through a lightweight Gated-Recurrent-Unit (GRU) network that aids detection without severe overheads. Traditionally, attention operations perform input compression using convex combinations where the weights are determined using neural networks. 

\textbf{Generative Models.} More recent works such as USAD~\cite{audibert2020usad},  MAD-GAN~\cite{mad_gan} and openGauss~\cite{li2021opengauss} do not use resource-hungry recurrent models, but only attention-based network architectures to improve training speeds. The USAD method uses an autoencoder with two decoders with an adversarial game-style training framework. This is one of the first works that focus on low overheads by using a simple autoencoder and can achieve a several-fold reduction in training times compared to the prior art. The MAD-GAN~\cite{mad_gan} uses an LSTM based GAN model to model the time-series distribution using generators. This work uses not only the prediction error, but also the discriminator loss in the anomaly scores. The openGauss approach uses a tree-based LSTM that has lower memory and computational footprint and allows capturing temporal trends even with noisy data. However, due to the small window as an input and the use of simple or no recurrent models, the latest models are unable to capture long-term dependencies effectively. A recently proposed HitAnomaly~\cite{huang2020hitanomaly} method uses vanilla transformers as encoder-decoder networks, but is only applicable to natural-language log data and not appropriate for generic continuous time-series data as inputs. Other methods, such as TopoMAD~\cite{topomad}, use a topology-aware neural network that is composed of a Long-Short-Term-Memory (LSTM) and a variational autoencoder (VAE) to detect faults. \blue{However, the reconstruction error is only obtained for the last state, limiting them to using reactive fault recovery policies. Similar methods use slight variations of LSTM networks with either dropout layers~\cite{girish2021anomaly}, causal Bayesian networks~\cite{gan2020sage} or recurrent autoencoders~\cite{chouliaras2021detecting}.} A GAN-based approach that uses a stepwise training process, StepGAN~\cite{stepgan}, converts the input time-series into matrices and executes convolution operations to capture temporal trends. However, such techniques are not agnostic to the number of hosts or workloads as they assume a maximum limit of the active tasks in the system. Moreover, even though more accurate than heuristic-based approaches, deep learning models such as deep autoencoders, GANs and recurrent networks are adaptive and accurate, but have a high memory footprint that adversely affects system performance. \blue{To resolve this, some works have been proposed that have low memory footprint, such as GONs~\cite{tuli2021generative}. }

\textbf{Clustering Models.} Very recent works also propose a few-shot learning method for fault detection~\cite{won2021performance}. Other recent methods utilize deep neural networks to execute fuzzy clustering~\cite{awgg, li2018comparison}. For instance, the Adaptive Weighted Gath-Geva (AWGG)~\cite{awgg} clustering method is an unsupervised model that detects faults using stacked sparse autoencoders to reduce detection times. Such methods train using supervised labels and do not present a mechanism to recover from faults once detected, \blue{and hence can not be used to develop end-to-end fault tolerance in Fog systems.} Other methods, such as Isolation Forest~\cite{liu2008isolation} in an unsupervised method that is used for anomaly detection in systems~\cite{tuli2022tranad}.

\subsubsection{AI Augmented Fault Remediation}
\label{sec:maintenance_remediation}
When Edge servers fail or are unavailable, optimal migration of the running tasks is crucial. However, it is also important to ensure that the task placement and scheduling procedures are fault-aware and aim to minimize system faults to minimize the overheads of running remediation strategies. 

\textbf{Fault-Aware Scheduling.} Recently, several resilience models have been proposed that leverage AI methods like RL, surrogate or reconstruction modeling. Many of these methods run proactive scheduling and task placement steps to avoid faults in a future state. An RL based approach is Load Balancing and Optimization Strategy (LBOS)~\cite{lbos} that allocates the resources using RL. The reward of the RL agent is calculated as a weighted average of multiple QoS metrics to avoid system contention by balancing the load across multiple compute nodes. The values of the weights are determined using genetic algorithms. LBOS observes the network traffic constantly, gathers the statistics about the load on each Edge server, manages the arriving user requests and uses dynamic resource allocation to assign them to available Edge nodes. However, RL approaches are known to be slow to adapt in dynamic settings~\cite{tuli2021cosco}. Most other approaches use neural networks as a surrogate model. For instance, Effective Load Balancing Strategy (ELBS)~\cite{elbs} is a recent framework that offers an execution environment for IoT applications and creates an interconnect among Cloud and Edge servers. The ELBS method uses the priority scores to proactively allocate tasks to Edge nodes or worker nodes as brokers to avoid system failures. It uses a fuzzy inference system to calculate the priority scores of different tasks based on three fuzzy inputs: SLA deadline, user-defined priority, and estimated task processing time. The priority values are generated by a neural network acting in the capacity of a surrogate of QoS scores. The Proactive Coordinated Fault Tolerance (PCFT)~\cite{pcft} method uses Particle Swarm Optimization (PSO) to reduce the overall transmission overhead, network consumption and total execution time for a set of tasks. This method first predicts faults in the running host machines by anticipating resource deterioration and uses PSO to find target hosts for preemptive migration decisions. This approach mainly focuses on reducing transmission overheads in distributed Cloud setups but often fails to improve the I/O performance of the compute nodes. CSAVM~\cite{satpathy2018crow} uses another evolutionary Crow Search scheme to take live migration decisions for the task queues. The method is used to optimize the power consumption of a compute setup by preventing unnecessary migrations. DDQP~\cite{wang2021ddqp} uses double deep Q-networks to place services on network nodes. \blue{However, such reinforcement learning schemes are known to be slow to adapt in volatile settings~\cite{tuli2021cosco}.} Another such work is PreGAN~\cite{tuli2021pregan}, which uses a GAN to generate preemptive migration decisions and anomaly scores from an input Fog system state. It uses a co-simulator in tandem with a few-shot anomaly classifier to ensure robust model training and fine-tunes the model to adapt to dynamic scenarios. 

\textbf{Load Balancing for Tolerance.} Load balancing is a concept that proactively aims to balance the load on different elements of a Fog infrastructure to avoid a faulty future system state. One such work, namely DPSO~\cite{jan2021ai}, proposes network gateways to host the load balancing logic where they monitor the load across Edge servers and balance the load using evolutionary algorithms. Furthermore, a migration mechanism is also incorporated where application modules are rearranged to achieve a balanced load across Edge servers. Migration is triggered based on a machine learning-based dynamic threshold. Similarly, FOCALB~\cite{kaur2021focalb} proposes a hybrid load balancing algorithm based on Grey Wolf Optimization (GWO) and Ant Colony Optimization (ACO), where energy consumption, execution time and implementation cost of scientific workflows are achieved by uniformly distributing the workload across Fog devices to optimize the Fog resource utilization. Similarly, some methdos~\cite{arabnejad2017fuzzy} propose a fuzzy logic based weighting scheme to run load-balancing task placement. Other methods, like PEFS~\cite{marahatta2020pefs}, present a prediction-based energy aware fault-tolerant load balancing scheme that uses a neural network to predict faults in the system and run load-balancing strategies to ensure a high resource utilization ratio. \blue{Load balancing based approaches are proactive in terms of fault-tolerance and do not require to use additional compute infrastructure in case of failures, making them more suitable for resource constrained settings.}

\textbf{Automatic Scaling.} Many methods aim to optimal decide how to scale the Fog infrastructure to avoid or recover from faults in the system. Similar to resource provisioning (Section~\ref{sec:deployement_provisioning}), here too, it is vital to avoid over-provisioning and under-provisioning of limited Fog resources under dynamic workloads. An RNN based method~\cite{etemadi2021cost} provides a deep learning based solution to utilize metrics such as resource requests (\textit{i.e.}, CPU, RAM, etc.) and Fog resource status (e.g., CPU efficiency, storage utilization, network traffic, active/inactive resources) to make optimum auto-scaling decisions. Moreover, AI-augmented auto-scaling methods have the potential to support proactive auto-scaling of containers under dynamic workload fluctuations. Similarly, \cite{abdullah2020predictive} presents a predictive auto-scaling policy using decision tree regression (DTR) model where a reactive rule-based auto-scaling mechanism is employed to train the proactive model under multiple heterogeneous workloads. Existing works explore the use of AI-augmented workload forecasting~\cite{li2020heterogeneity, etemadi2020learning} to make proactive auto-scaling decisions within Edge/ Fog environments. Another method, MADRP~\cite{li2020heterogeneity}, uses a hybrid ARIMA and DNN model to forecast the workloads, whereas a nonlinear autoregressive (NAR) neural network is used by \cite{etemadi2020learning} to predict the future demands for the Fog devices. Methods such as FLBFH~\cite{DBLP:journals/corr/abs-2103-06381}, propose a fuzzy logic-based method to handle unpredicted and predicted failures in Fog environments. Such methods predict two failure scores to decide what actions to be undertaken to handle failures for unreliable Fog devices. The first failure score is based on device mobility, device response time and device power availability. This score determines the checkpointing interval as a proactive mechanism for unpredicted failures. The second score is based on CPU utilization, device mobility, device response time, device power availability, device communication. This score is used to decide about preemptive task migration. In some cases where the rate of unpredicted failure is high, the proposed fuzzy-logic mechanism will suggest an application replication. An extension to this work is the Dependency and Topology-aware Failure Resilience (DTFR) algorithm, which considers failure probability, response time and the number of replicas to schedule services on Edge servers~\cite{9303420}. DFTR explores the spatio-temporal failure dependency among Edge servers to develop a dynamic method with minimum redundancy to enhance the failure resilience of services. %It uses a machine learning based method to compute the join failure probability of Edge servers. The failure dependencies are modeled as a dynamic Bayesian network that is trained from the historical traces. Combining the joint failure probability and the link failure probability will provide the overall service failure probability. The replication mechanism is adopted to schedule services on nearby servers to improve service availability. 
Other works~\cite{aral2017quality}, continue this trend to estimate the availability level of VMs in Edge Data Centers (EDCs) based on Bayesian Networks. The probabilistic models consider dependencies between different failures such as hardware, software, or network failures, and power outage. This model is utilized to select VMs that can meet the availability requirements in SLA. Another similar method is the Fuzzy-based Real-Time Auto-scaling (FRAS)~\cite{fras} technique that leverages a virtualized environment for the recovery of IoT applications that run on compromised or faulty Edge nodes. Here, FRAS executes each IoT application in a virtual machine (VM) and performs VM autoscaling to improve execution speed and reduce execution costs. The VM autoscaling decisions making involves inference of system QoS using a fuzzy recurrent neural network as a surrogate model.

\begin{table*}[t]
    \centering
    \caption{Classification of state-of-the-art techniques in terms of the used AI methods.}
    \label{tab:trends}
    \resizebox{\linewidth}{!}{
\begin{tabular}{|>{\raggedright}p{3cm}|>{\raggedright}m{4.4cm}|>{\centering}p{2.5cm}|>{\centering}p{2cm}|>{\centering}p{2cm}|>{\centering}p{2cm}|>{\centering}p{2cm}|>{\centering}p{2cm}|}
\hline 
\multirow{2}{3cm}{Category} & \multirow{2}{4cm}{Method} & \multicolumn{2}{c|}{Deployment} & \multicolumn{2}{c|}{Scheduling} & \multicolumn{2}{c|}{Maintenance}\tabularnewline
\cline{3-8} \cline{4-8} \cline{5-8} \cline{6-8} \cline{7-8} \cline{8-8} 
 &  & DNN Deployment & Resource Provisioning & Bag-of-Tasks & Workflows & Detection and Prediction & Tolerance\tabularnewline
\hline 
\hline 
\multirow{5}{3cm}{Classical AI} & \cellcolor{yellow!25}Informed and Local Search & \cite{jiang2019model, cnn_slimming, zhou2021bbnet, sattler2019robust, luo2021novel, lane2016deepx, imani2019floatpim, kang2017neurosurgeon, zhang2021deepslicing, kaplunovich2020automatic} & . & \cite{panda2015uncertainty, jawad2018robust, alelaiwi2019efficient} & \cite{alsurdeh2021hybrid} & . & \cite{satpathy2018crow, marahatta2020pefs}\tabularnewline
\cline{2-8} \cline{3-8} \cline{4-8} \cline{5-8} \cline{6-8} \cline{7-8} \cline{8-8} 
 & \cellcolor{orange!25}Neural Design & \cite{shao2020bottlenet++, yu2020easiedge, Huang2020, reisizadeh2020fedpaq, hamer2020fedboost, tran2019federated, coelho2021automatic, jain2018compensated, matsubara2019distilled, teerapittayanon2017distributed, goli2020migrating, chen2018thriftyedge, huang2020clio, kim2017splitnet} & . & . & . & \cite{tuli2021generative} & .\tabularnewline
\cline{2-8} \cline{3-8} \cline{4-8} \cline{5-8} \cline{6-8} \cline{7-8} \cline{8-8} 
 & \cellcolor{pink!25}Maxweight, Linear/Dynamic Prog. & \cite{callegaro2020optimal} & \cite{semidirect} & \cite{liu2020pond} & \cite{wan2012qos} & . & .\tabularnewline
\cline{2-8} \cline{3-8} \cline{4-8} \cline{5-8} \cline{6-8} \cline{7-8} \cline{8-8} 
 & \cellcolor{red!25}Regression (Linear/Gaussian/SVM) & . & \cite{hyndman2018forecasting, zhu2016pso, chen2018resource, uahs, cahs, taylor2018forecasting, arima, calheiros2014workload, xu2020hybrid} & \cite{jamshidi2016uncertainty, bui2017energy, onggo2021combining, onggo2018symbiotic} & \cite{wang2020intelligent} & \cite{dagmm} & \cite{abdullah2020predictive, li2020heterogeneity}\tabularnewline
\cline{2-8} \cline{3-8} \cline{4-8} \cline{5-8} \cline{6-8} \cline{7-8} \cline{8-8} 
 & \cellcolor{magenta!25}ACO/AMO/PSO/GA/BO/GWO & . & \cite{aco, uahs, cahs, zhu2016pso, chen2020self, asghari2021task} & \cite{han2018fog, tuli2020ithermofog} & \cite{impso, huang2020ant, ghanavati2020energy, dnsga, esvr} & . & \cite{pcft, jan2021ai, kaur2021focalb}\tabularnewline
\hline 
\multirow{3}{3cm}{Reinforcement Learning} & \cellcolor{cyan!25}Tabular RL (SARSA/Q Learning) & . & . & \cite{wang2021energy} & . & . & \cite{lbos}\tabularnewline
\cline{2-8} \cline{3-8} \cline{4-8} \cline{5-8} \cline{6-8} \cline{7-8} \cline{8-8} 
 & \cellcolor{blue!25}Deep RL (DQN) & \cite{tuli2021splitplace} & \cite{xu2020recarl, bitsakos2018derp, sami2021ai} & \cite{tang2018migration, li2019deep} & \cite{wang2019multi, kaur2020deep} & . & \cite{wang2021ddqp}\tabularnewline
\cline{2-8} \cline{3-8} \cline{4-8} \cline{5-8} \cline{6-8} \cline{7-8} \cline{8-8} 
 & \cellcolor{Fuchsia!25}Policy Gradient Learning & \cite{xia2019dnntune, li2019edge, zhao2021edgeml, gillis} & \cite{xu2020deep, chen2020deep, chen2021deep} & \cite{sheng2021deep, tuli2020dynamic, cheng2021multi} & \cite{ghosal2020deep, hu2019learning} & . & .\tabularnewline
\hline 
Neural Optimization & \cellcolor{green!25}Deep Surrogate Optimization & . & \cite{decisionnn} & \cite{tuli2021cosco, tuli2021hunter, tuli2021gosh} & \cite{tuli2021mcds} & . & \cite{elbs}\tabularnewline
\hline 
\multirow{4}{3cm}{Neural Approximation} & \cellcolor{LimeGreen!25}Fully Connected Network (FCN) & . & \cite{narya, chen2019iraf} & \cite{tuli2021cosco, tuli2021gosh} & . & . & .\tabularnewline
\cline{2-8} \cline{3-8} \cline{4-8} \cline{5-8} \cline{6-8} \cline{7-8} \cline{8-8} 
 & \cellcolor{SeaGreen!25}Convolutional Neural Network (CNN) & . & \cite{bega2019deepcog, jeddi2019water} & . & . & \cite{mscred, cae_m} & .\tabularnewline
\cline{2-8} \cline{3-8} \cline{4-8} \cline{5-8} \cline{6-8} \cline{7-8} \cline{8-8} 
 & \cellcolor{OliveGreen!25}Recurrent Neural Network (GRU/LSTM) & . & \cite{lstm, yazdanian2021e2lg} & . & \cite{feng2019content} & \cite{lstm_ndt, omnianomaly, lstm_vae, topomad, girish2021anomaly, chouliaras2021detecting} & \cite{etemadi2021cost, tuli2021start}\tabularnewline
\cline{2-8} \cline{3-8} \cline{4-8} \cline{5-8} \cline{6-8} \cline{7-8} \cline{8-8} 
 & \cellcolor{gray!25}GAN/GNN/Transformers/Fuzzy-Logic & . & . & \cite{tuli2021pregan, guan2021gan} & . & \cite{mtad_gat, gdn, mad_gan, stepgan, won2021performance, li2018comparison, awgg} & \cite{tuli2021pregan, arabnejad2017fuzzy, DBLP:journals/corr/abs-2103-06381, 9303420, aral2017quality, 8567669}\tabularnewline
\hline 
\end{tabular}}
\end{table*}

\textbf{Straggler Aware Models.} Another common performance problem in Fog systems is dealing with straggler tasks that are slow running instances that increase the overall response time. Such tasks can significantly impact the system's QoS and the SLA. Methods, such as JFP~\cite{8567669}, exploit failure dependencies between Edge servers to predict the failure probability of given service deployment. JFP evaluates the use of replication in Edge servers based on analyzing historical failure logs of individual servers, modeling temporal dependencies as a Dynamic Bayesian Network (DBN), and predicting the probability at which a certain number of servers fail simultaneously. It also uses two replica scheduling algorithms to optimize failure probability and the cost of redundancy in an Edge computing environment. \blue{Similarly, other methods such as START~\cite{tuli2021start}, proactively predict the occurrence of straggler tasks to avoid adverse impact on system QoS. }

%%%%%%%%%%%%%%%%% METHOD OVERLAP ACROSS SOTA %%%%%%%%%%%%%%%%% 

\section{Classification of State-Of-The-Art}
\label{sec:classes}

Table~\ref{tab:trends} classifies the discussed state-of-the-art works in Section~\ref{sec:sota} as per the AI methods they use. This facilitates researchers in identifying the class of methods that have been used in the past and can be utilized to solve one of the scopes of deployment, scheduling and maintenance for resource management in Fog systems.

\textbf{Classical AI.} This category includes traditional AI schemes that do not utilize DNNs, such as local and evolutionary search, regression and meta-heuristic optimization schemes. We also include neural design, \textit{i.e.}, the application specific design of neural models to achieve optimal performance or reduced overheads. We observe that the search and design based methods are quite popular in the case of DNN deployment. This is predominantly due to the search-driven DNN design specific improvements required to ensure optimal deployment of large-scale AI models on constrained Fog nodes. Nevertheless, we see some overlap across the three domains: deployment, scheduling and maintenance. For instance, risk-based and robust optimization has been seen in gradient pruning for DNN deployment~\cite{sattler2019robust}, task scheduling~\cite{jawad2018robust, panda2015uncertainty} and load balancing~\cite{marahatta2020pefs}. Similarly, neural design has been used for splitting DNNs for inference of resource-constrained Edge nodes~\cite{kim2017splitnet} and memory-efficient anomaly detection~\cite{tuli2021generative}. Regression models have been popular across all domains. All kinds of predictions are made using regression techniques, such as workload demand prediction for optimal resource provisioning~\cite{zhu2016pso, uahs, cahs}, task QoS prediction~\cite{jamshidi2016uncertainty, tuli2020ithermofog} and time-series reconstruction for anomaly detection~\cite{dagmm}. Similarly, meta-heuristic optimization strategies have frequently been in use in Fog research. For instance, PSO optimization has been used for decision optimization of VM provisioning in Cloud~\cite{zhu2016pso, chen2020self}, workflow scheduling decision~\cite{impso} and load-balancing~\cite{jan2021ai}. 

\textbf{Reinforcement Learning.} This category includes the various ways to solve MDP style problems using AI methods, such as tabular RL (Q and SARSA learning), deep Q learning and policy gradient methods (A3C, DDPG, etc.). Most state-of-the-art approaches do not utilize tabular RL due to its poor scalability of modeling real-life state-action spaces in physical Fog and Cloud systems with thousands of devices. Thus, researchers tend to rely on neural network-based approximation of the Q function, which estimates the long-term reward using a DNN. DQNs have been used to optimize the placement decisions of neural network-based tasks generated after DNN splitting~\cite{tuli2021splitplace}, QoS aware resource provisioning~\cite{bitsakos2018derp}, task and workflow scheduling~\cite{tang2018migration, li2019deep, wang2019multi} and fault-tolerant scheduling~\cite{wang2021ddqp}. However, in complex Fog scenarios, a richer action space might be required to ensure that RL agents do not get stuck in local optima. As tabular RL and DQNs have deterministic action policies, researchers have shifted to utilizing DNNs for stochastic action prediction, \textit{i.e.}, policy gradient learning (PGL). These include REINFORCE, Actor-Critic and other forms of DNNs that predict action probabilities instead of Q values. For instance, some methods use PGL to decide the optimal placement of split neural models in a heterogeneous Edge-cloud setup~\cite{gillis}. Other methods utilize PGL to optimize metrics such as energy~\cite{tuli2020dynamic} and cost~\cite{cheng2021multi} by using them as reward signals. 

\textbf{Neural Optimization and Approximation.} We consider another category of optimization that either uses DNNs to approximate optimization objects (unlike regression in classical AI) or gradient-optimization to generate optimal decisions (unlike ACO, PSO, etc. in classical AI). Recently, many methods use DNNs as QoS surrogates, to optimize provisioning decision~\cite{decisionnn}, scheduling decisions~\cite{tuli2021cosco, tuli2021gosh, tuli2021mcds} or fault-tolerant scheduling~\cite{elbs}. Other methods utilize DNNs directly to take deployment, scheduling or maintenance decisions. For instance, FCNs~\cite{narya, chen2019iraf}, CNNs~\cite{bega2019deepcog, jeddi2019water} and LSTMs~\cite{lstm, yazdanian2021e2lg} are used for resource provisioning, FCN~\cite{tuli2021cosco} and LSTM~\cite{feng2019content} are used for scheduling, and CNNs~\cite{mscred, cae_m} and RNNs~\cite{lstm_ndt, omnianomaly, lstm_vae, topomad, girish2021anomaly, chouliaras2021detecting} are also used for reconstruction based fault detection. Fog maintenance related state-of-the-art methods also leverage other DNN types, including GNNs and GANs~\cite{tuli2021pregan, arabnejad2017fuzzy, DBLP:journals/corr/abs-2103-06381, 9303420, aral2017quality, 8567669}.

% Other methods rely on co-simulators to generate labelled data in a self-supervised fashion~\cite{deepft}.

%%%%%%%%%%%%%%%%% FUTURE DIRECTIONS %%%%%%%%%%%%%%%%% 

\section{Trends, Challenges and Future Directions}
\label{sec:future_directions}
% how can a method in one domain be adapted for other domains
Existing AI-driven resource management techniques cover a wide range of decision making problems. We now identify the key trends in the domain of AI based augmentation for resource management in the Fog continuum and elucidate them in Section~\ref{sec:trends}. We also discuss in Section~\ref{sec:limitation}, the limitations of the current state-of-the-art works as per the classes identified in Section~\ref{sec:classes}. Stemming from the identified limitations, we discuss emerging challenges in the field of AI-augmented Fog continuum systems and a series of open opportunities while briefly proposing new methods for future blue-sky research in application areas (Section~\ref{sec:application_areas}) and AI methods (Section~\ref{sec:methods}).

% A major drawback of such surrogate modeling methods is that their parameters need to be periodically fine-tuned to adapt to dynamic environments, giving rise to high overheads -> confidence aware. 
% co-simulations  -> even higher overheads. need different levels of co-simulations.
\subsection{Trends}
\label{sec:trends}
\textbf{Shift to Deep Surrogate Models.} Recently, there has been a shift from using regression models, such as linear regression, support vector regression, Gaussian process regression, to training a DNN. Regression models allow us to tune the parameters of a function using expectation maximization~\cite{aima}. These models are typically used to generate an estimate of the system performance, usually a combination of QoS metrics, with respect to independent variables like resource management decision~\cite{cahs, uahs}. However, in practice, the data distributions that these models try to capture may have far more complex relationships with independent variables that such models can represent. To combat this, researchers now resort to DNNs as function approximators and surrogates of QoS metrics such as energy consumption, average response time and execution costs~\cite{decisionnn, tuli2021cosco, tuli2021hunter}.

\textbf{Shift to Co-Simulated Digital-Twins.} AI models that rely on DNNs for function approximation, such as deep surrogate optimization, DQN, and policy gradient methods often face present issues characteristic of DNNs. For instance, when training a DNN to predict workload demands in a future state, it is often trained with historical trace data collected from a Fog system. However, when the model is applied in a setting with different workload traces, the model has poor demand prediction accuracy as it is never exposed to new data at training time. This problem is commonly referred to as the \textit{exposure bias} problem in DNN training. To tackle this, recent methods now develop a co-simulated digital twin of the Fog system to solve three problems~\cite{tuli2021cosco, elbs}. First, for data augmentation, \textit{i.e.}, generate new traces by random perturbation of the environment or workload parameters to solve the exposure bias problem. Second, to solve the \textit{data saturation} problem, \textit{i.e.}, increasing the amount of data does not improve the model performance. Co-simulation allows us to run A/B tests to generate diverse scenarios to improve model performance. Third, co-simulations allow the generation of new datapoint for the latest system state, facilitating fine-tuning DNNs to adapt to non-stationary workload settings.

\textbf{Shift to Transformers and Geometric Models.} For series like data, researchers traditionally used recurrent models such as GRUs and LSTMs. However, training these models is time-consuming, giving rise to high training costs on public Cloud or local Edge nodes. The main bottleneck of such models is the requirement of providing sequential data one at a time~\cite{vaswani2017attention, tuli2022tranad}. Recent models, like Transformers, use self-attention to infer on the complete sequence at once, allowing faster training and higher accuracy. Further, instead of FCN, CNNs and LSTMs, researchers are now resorting to composite AI neural networks that also model the system state as a geometric model, most often as a graph. These might be used to encode the network architecture~\cite{gdn} or the input decision~\cite{tuli2021hunter}. GNNs over graph-like data allow capturing data correlations to be aware of the spatial structures.

\textbf{Shift to Resource Efficient Management.} Most AI-based resource management applications are heavy in terms of resource requirements. Thus, broker nodes are typically more powerful than a common Fog worker~\cite{tuli2019fogbus, gill2019transformative}. As the number of devices in the worker layer of Fog architectures increases, the resource management AI models become more data and resource hungry. To scale and allow resilience in the broker layer of Fog systems, running resource management applications on worker-like resource-limited devices becomes inevitable. Thus, systems based AI research is now working to develop more pragmatic AI models that can be deployed in decentralized and constrained environments~\cite{tuli2021generative, li2020federated, slimgan}. Further, researchers are developing DNNs that have much lower training times than before, facilitating quick adaptability in volatile environments~\cite{huang2020hitanomaly}. Another important trends is taking into account the energy consumption of DNNs while achieving high accuracy to have more sustainable machine learning models~\cite{tuli2021hunter}.

\textbf{Shift to Unsupervised Models.} Traditional methods mainly rely on manual labeling of important data characteristics, such as fault indication, through domain experts, which is infeasible in modern IoT solutions with enormous amounts of log data~\cite{awgg}. Thus, for large-scale systems, researchers are now developing unsupervised and semi-supervised models that are as accurate as supervised models~\cite{awgg, omnianomaly}. The advantage of unsupervised models is that we do not require labeled data, allowing us to scale resource management systems to systems with possibly millions of IoT devices and Fog nodes.

\subsection{Limitations}
\label{sec:limitation}
\textbf{Scalability.} Most AI models suffer from the limitation of having poor scalability. Scalability in Fog systems refers to the ability to apply an AI model as the number of Fog nodes or workloads increases without a significant drop in system performance. As the number of IoT devices and users relying on Fog architectures increases, it makes it crucial to develop scalable AI models. This specifically requires developing on top of existing AI methods that are scalable. For instance, tabular reinforcement learning (Q/SARSA learning) saves the Q estimate of each state and action pair in an MDP formulation. On the contrary, DQN utilizes a DNN to capture the interdependence across states and actions to train a generic model that can provide accurate Q estimates without needing the same number of parameters as tabular RL methods. Similarly, other techniques such as geometric modeling (GNNs) and attention operation further improve scalability.  

\textbf{Generalizability.} Generalizability is the ability of an AI model to perform successfully for unseen inputs. This depends on the stability, robustness and adaptability of the developed model~\cite{xu2012robustness}. Thus, limitations in the above areas result in limiting the generalizability of the model. To improve the generalizability of the traditional machine learning and deep learning algorithms within the context of distributed architectures such as Edge and Fog computing, federated learning is a viable option. It allows the models to be trained with a sufficient amount of data when the collection of data at a central location is not a possible option due to privacy, security or the sheer volume of data generated. However, conventional federated learning needs to be further improved to adapt to the challenges of network, computation and storage heterogeneity within Edge/ Fog environments \cite{hosseinalipour2020federated}.

\textbf{Reliability.} Reliability limitations of AI techniques can be analyzed under two main aspects: stability and robustness. The former indicates the ability of a model to yield consistent performance across similar yet diverse data inputs. The latter indicates the consistency of the output of the approach under new data~\cite{xu2012robustness}. IoT data-related issues such as missing data due to unreliable networks, limited access to sensitive data (for example, health data), noisy data and malicious data, cripple the stability and robustness of the machine learning and deep learning approaches. Susceptibility of machine learning models, especially deep learning models to adversarial examples, is critical within the context of latency and safety-critical IoT applications where accuracy is paramount \cite{qiu2021special}. Meta-heuristic algorithms also face limitations in stability and robustness. As meta-heuristics are designed to converge towards a near-optimal solution, stability and robustness limitations occur due to their tendency to converge to local optimum solutions, especially due to the dynamic changes in Fog environments.

\textbf{Adaptability.} Adaptability is the ability of the AI models to maintain accuracy when training and test data belong to different distributions. However, traditional machine learning and deep learning approaches operate under the assumption that both training and test data share the same distribution, which results in performance reduction in real-world deployments, for instance, in cases with exposure bias~\cite{pan2009survey}. Insufficient and biased data (\textit{i.e.}, due to data privacy and security issues in smart healthcare, IIoT, etc.), outdated training data and inability to use large data sets in Fog environments due to resource limitations demand higher adaptability in such use cases \cite{sufian2020survey}. To overcome this limitation, some machine learning and deep learning approaches leverage transfer learning, which is a learning framework that enables knowledge transfer between task domains~\cite{pan2009survey}. Some AI models utilize co-simulators to generate diverse datapoints and adapt to changing settings~\cite{tuli2021cosco}. However, fine-tuning the parameters of DNNs using co-simulators gives rise to high overheads. 

\textbf{Agility.} Agility indicates the ability of a system to adapt and evolve rapidly with changing Fog environments. This becomes a prominent requirement in IoT applications, Fog environments that keep evolving rapidly require high agility not only within application development and deployment but also for algorithm development for resource provisioning, application scheduling and system maintenance. To keep up with this nature, AI models used within these contexts need to be able to undergo rapid updates as more data and data sources appear, more service requirements appear and the nature of the deployment environment and its technologies evolve (i.e., updates in communication technologies, availability of novel Edge/ Fog computation resources and their hardware or architectural changes etc.) \cite{jackson2019agile}. But, the data-centric nature of the lifecycle of an AI model makes the development, testing, deployment cycle highly experimental and repetitive, thus making agility a major limitation~\cite{schleier2015architecture}.

\subsection{Emerging Challenges}
\textbf{Legacy Deployment.} As research progresses and more accurate and better performing models are developed, it typically follows the adoption of advanced AI models by industry. However, more complex AI models are usually more resource-hungry and need more powerful systems to be deployed on. To deploy an enhanced AI model, technology based companies such as Meta, Amazon, Netflix and Google frequently upgrade their devices, raising many sustainability concerns~\cite{gill2019transformative}. Stemming from the scalability limitations of state-of-the-art, the integration of large-scale DNN models within legacy edge or cloud machines has become a challenging problem. As research moves in the direction of the neural design of sophisticated AI models, it becomes crucial to ensure that these new models can be deployed on legacy infrastructures to bring down deployment costs and carbon footprint of AI.

\textbf{Automated and Generic Modelling.} Another challenge being faced by industrial AI adopters, related to the generalizability and adaptability of AI models, is the ability to tune AI models in settings different from the ones tested by researchers~\cite{liaw2018tune}. As the performance of AI models is highly dependent on the proper tuning of a large number of hyperparameters, these variables need to be re-tuned when deploying a pre-trained model in a new setting of scheduling or fault detection in Edge/Cloud. This problem stems from generalizability, but needs to be solved specifically for each application domain of deployment, scheduling or maintenance. In such cases, either the hyperparameter values of the models need to be decided in an automated fashion, or the neural design needs to be generic enough to accommodate new Fog settings, with possibly different number of nodes, workload characteristics and user demands.

\textbf{Interpretability.} Many state-of-the-art AI methods are being utilized today as black-box models that give rise to high QoS in Fog systems, but do not have any transparency on the process that led to the various resource management level decisions made by an AI agent. This mainly entails explaining the main reasons for choosing or not choosing certain management decisions and exploring the unknown state spaces to ensure exhaustive coverage of the decision space. For sensitive industrial segments, such as healthcare and autonomous vehicles, it is crucial to expose the underlying patterns and features in the decision-making process to gain credibility for the end-user. Building such white-box or explainable models is an emerging field of research for the use of trustworthy AI models. Many AI models, such as decision trees, regression algorithms and rule-based systems allow interpretability, but are not as accurate or scalable as deep-learning based counterparts.

\subsection{Application Areas}
\label{sec:application_areas}
\textbf{Healthcare.} With the rapid increase in connected devices in hospitals, such as sensors, mobile phones and wearables, the amount of data generated and their rate of generation is snowballing. This results in a massive increase in the volume and variety of available health data, paving the way for the development of more reliable and robust AI models in the areas of proactive monitoring, disease prevention, and more in smart healthcare~\cite{panesar2019machine}. However, this results in challenges related to ensuring data quality and security, especially in the context of distributed EdgeAI, particularly in the case of handling sensitive healthcare information like patient records. To overcome these challenges, future research is focusing on the convergence of Blockchain and AI, where Blockchain is used in solving data quality and integration issues that enable AI to improve the accuracy of data analytics~\cite{yaqoob2021blockchain}. Moreover, low latency communication technologies like 5G/6G enable novel technologies like Augmented Reality, Virtual Reality and Tactile Internet, thus improving and expanding services such as robot-assisted surgery.  

\textbf{Next Generation Networking.} Due to the high data rate, reliability, ultra-low latency, and ultra-low energy consumption provided by 5G/6G, these wireless communication technologies are identified as the key enablers of future IoT applications. However, to support the ever-evolving service requirements of the IoT services, 5G/6G technologies have to be able to observe environment variations and dynamically self organize the network accordingly \cite{li2017intelligent}. This can be achieved through AI-empowered management and orchestration of cellular resources within 5G/6G networks. With the advancements in Software Defined Networking (SDN) and network function virtualization (NFV), 5G/6G technologies introduce Network Slicing (NS) to support this. NS is a mechanism for provisioning virtualized network resources intelligently based on the service performance metrics \cite{letaief2019roadmap}. Learning algorithms such as deep learning and reinforcement learning can improve the dynamism of NS through prediction-based proactive resource allocation in Edge or Cloud architectures with dynamic slice creations for different applications \cite{wijethilaka2021survey}. \blue{We also need to consider upcoming cases where the internet is provided through satellites (such as the Starlink network) in lieu of the conventional copper/fibre connections. In such cases, the latency and bandwidth characteristics might be significantly different, requiring re-tuning hyperparameters or adapting existing AI based resource management policies~\cite{song2021energy,wang2019satellite}.}

\textbf{Production and Supply Chain.} In the coming age of IIoT, most industrial pipelines are managed by smart devices. Such devices may utilize AI methods to self-monitor and predict potential problems in the supply chain to optimize the overall service efficiency. The COVID-19 pandemic is an example demonstrating the importance of automation in logistics to avoid service downtimes~\cite{salehi2021relief}. AI based forecasting approaches, such as recurrent neural models and GANs can be used to predict stock shortage and proactively order additional stocks to prevent shortage~\cite{salehi2021relief}. Similarly, large-scale ML models can aid the development of smart-manufacturing technologies that utilize several IoT and Fog devices to collaboratively monitor, control manufacturing and production related equipment.

\textbf{Smart Cities.} Smart cities aim to utilize IoT to deliver services that can enhance the living standards within cities. This includes a plethora of application domains such as smart governance, smart energy, smart transportation, and smart security~\cite{nayak2021intelligent}. Advancements in wireless communications such as 5G/6G enable a massive amount of data to be transmitted towards Edge nodes in real-time. This has resulted in the rise of novel technologies like crowd-sensing and crowd-sourcing~\cite{kong2019mobile}. Distributed collections and processing of such massive volumes of data demands future research to focus on distributed and reliable AI, specifically in data-sensitive applications at the Edge. At the same time, data security becomes crucial in future smart city services, with the widespread use of crowd-sensing and crowd-sourcing for data collection. Moreover, ultra-low latency communication provided future radio access networks to support services related to hazard avoidance and safety~\cite{rudd2020repeatable}, which requires AI models with higher reliability, accuracy and lower latency.

\subsection{Methods}
\label{sec:methods}
\textbf{Self-supervised AI.} The self-supervised learning technique enables learning with unlabeled data by solving pretext tasks~\cite{saeed2020federated}. In contrast to this, supervised learning depends on the availability of labeled data. Even though a massive amount of data gets generated by the sensors, the lack of annotated data poses an obstacle for using supervised learning. This is specifically applicable in scenarios where fault-detection or workload scheduling is required for previously unseen workload or device characteristics in Edge or Cloud platforms. As generating expert-labeled fault labels or optimal scheduling decisions is infeasible for large-scale systems, self-supervised learning offers a possible solution for such scenarios. This approach is capable of generating a more generalizable model by removing the heavy dependency on labeled data by automatically generated data annotations, possibly using a co-simulator~\cite{saeed2020federated}. Moreover, self-supervised learning has the potential to achieve higher reliability due to its robustness to adversarial examples, label corruption, and input corruptions~\cite{self-supervised}. 

\textbf{Model Driven RL.} As Fog environments and service demands are dynamic; algorithms should have the capability to adapt accordingly. IoT applications (\textit{i.e.}, healthcare, smart cities, etc.) and their enabling telecommunication technologies (i.e., 5G, 6G) benefit from RL-based intelligence due to higher adaptability of RL techniques and their ability to learn without prior knowledge~\cite{sami2021ai}. However, exploration errors, long learning time and distributed learning within resource constraint devices in Edge environments are some of the challenges in utilizing RL techniques within resource-constrained and distributed Fog environments. A canonical case is the development of RL based methods for deploying large-scale application workflows on constrained Edge or Cloud clusters. Addressing these challenges in future research is vital for the RL-based approaches to reach their full potential within EdgeAI scenarios. In an attempt to address these challenges, EdgeAI research is exploring advanced RL approaches such as model-based RL~\cite{sutton2018reinforcement} and co-simulated RL~\cite{amini2020learning}.

\textbf{Analog AI.} The current implementation of AI is targeted for digital systems where the values are stored in a binary format. Herein, the major challenge posed by digital implementations of DNNs is the linear dependence of memory footprint with the number of parameters of the neural model. Upcoming analog memory-based chips present new ways to perform the same operations, but with orders of magnitude lower amount of memory requirement, computational load and energy consumed~\cite{channamadhavuni2021accelerating}. This is particularly useful in memory-constrained Edge and MEC devices where sophisticated DNNs need to be executed in \textit{AI on Fog} setting or decentralized resource management is required in \textit{AI for Fog} settings. However, there are some drawbacks in the loss of precision in computation across layers within a DNN. The tradeoff offered by such DNN implementations is similar to model pruning and splitting, but with possibly more extreme energy/compute benefits. This direction has been explored to a limited extent and requires further investigation and software development to efficiently harness the potential of analog hardware accelerators.

\textbf{Decentralized Modeling.} The success of distributed EdgeAI, by utilizing layer and semantic splitting strategies for AI deployment (discussed in Section~\ref{sec:sota}), shows some promise in other domains of scheduling and maintenance. For scheduling applications and maintaining Fog systems at scale, it is possible to decentralize the training and inference procedures of the resource management level AI applications across multiple broker nodes. The decentralized fashion of resource management has a two-fold benefit. Firstly, there is no single point of failure in the system as the management steps are run in multiple broker nodes. Secondly, it allows the distribution of resource management load across multiple computing devices, facilitating the scalability of the model. 

\textbf{AI Driven Simulations.} A major advantage of co-simulators in Fog systems is the ability to generate new data points for tuning AI models and resolving issues like exposure bias and data saturation~\cite{renda2020difftune}. However, another important benefit of co-simulators is the ability to run multiple simulations concurrently and pick the best resource management decisions, allowing interpretable decision-making. This is due to the ability of co-simulators to generate a complete execution trace, possible for several future states of the systems and allowing developers or end-users to visualize the long-term effects of various decisions. They are able to do this much faster than executing the decisions on a physical infrastructure and waiting for several minutes to reflect changes in QoS. This is primarily due to the discrete event-driven execution style of modern simulators. This is applicable for all three types of resource management decisions, \textit{i.e.}, deployment, scheduling and maintenance. Simulations can indicate changes in QoS scores for each model compression or splitting type, application placement or fault-remediation steps like preemptive migrations. These signals can facilitate decision making. However, co-simulators are mere approximations of the entire Fog system and typically fail to map the entire complexity of real infrastructure. However, the success of deep surrogate models hints us to build simulators with possibly millions of parameters and utilize DNNs to estimate optimal parameter values, such that our simulators resemble the real systems as closely as possible. An increased number of parameters could, in principle, give a higher representative capacity to our simulators, now being able to map complex real-life workloads and device characteristics. Thus, AI-driven simulators could help improve system performance with the added bonus of interpretable decision making.

\textbf{AI Driven Co-Design.} Currently, almost all resource management solutions for Fog systems solve only a specific problem from the three domains of deployment, scheduling or maintenance. However, for holistic performance enhancement, it is crucial to develop AI models that can concurrently take decisions across multiple facets of the management of Fog resources to efficiently exploit the synergy across these decision domains. Research in AI-based augmentation of Fog systems may benefit from other efforts in system co-design~\cite{hao2021enabling} to improve upon the existing management solutions. This is crucial in Fog systems particularly due to the constraints certain decision types impose on other resource management control knobs. For instance, provisioning decision constrains the devices on which incoming tasks can be scheduled on or the active tasks may be migrated to.  There is a need to build end-to-end AI models that rely on multi-modal data and can take multiple decision types simultaneously for data privacy and improved system performance.

\textbf{Foundation Models.} Recently, the AI community has demonstrated the broad impact and accuracy of large-scale foundation models~\cite{bommasani2021opportunities}, such as ChatGPT (based on GPT-4)~\cite{bubeck2023sparks}, LLaMA~\cite{touvron2023llama} and DALL-E~\cite{ramesh2022hierarchical}. These models are trained on broad data at scale and are adaptable to a wide range of downstream tasks. Such models can leverage large amounts of unlabeled data available throughout the web and open-source channels, to analyze broad level patterns in data that are applicable in most task-specific scenarios. Thus, it is possible to quickly fine-tune such models for a specific context and leverage the strong priors that such models provide. This has been made possible by the advent of Transformers and hardware optimizations~\cite{bommasani2021opportunities}.  Some recent works have also explored the applicability of language models such as Chat-GPT for task scheduling in cloud computing environments~\cite{wang2023chat}. As logs of resource utilization metrics can be represented as alphanumeric texts, these can be interpreted by such language models to provide insights on workload patterns. A quintessential example of leveraging such models is to fill in the hand-coded information in fog simulators and frameworks, such as resource capacities and functions of power consumption and other metrics of interest. Large-language models (LLMs) could be fine-tuned on datasets such as power consumption with CPU utilization of several processor types and could provide us with the relationship for a test processor given its characteristics such as core counts and clock speeds.  The main advantage of such models in the specific contexts of Edge and Fog environments is the new-found ability to zero-shot generalization to unseen settings. For instance, a model like LLaMa could possibly update the configurations and hyperparameters of a fog scheduler if we ask it to modify thresholds, heuristics and weights for a new environment with nodes having unseen resource capacities. Such directions may be pursued in the future for robust and reliable computing.

%%%%%%%%%%%%%%%%% CONCLUSIONS %%%%%%%%%%%%%%%%% 

\section{Conclusions}
\label{sec:conclusions}

This work conducts an extensive literature review of the methods concerned with AI-based augmentation of Fog systems. \blue{We discuss diverse state-of-the-art techniques for Fog resource management, specifically for optimal AI deployment, workload scheduling and system maintenance. We consider two kinds of AI models: AI \textit{on} and AI \textit{for} Fog computing. We use taxonomy of AI methods and classify them broadly into classical methods, machine learning, reinforcement learning and deep learning. There is significant overlap across different decision domains in terms of the used AI models.} This overlap suggests the importance of certain design decisions over others and hints at the possible gaps of current research. We have highlighted the importance of a more comprehensive research style that not only considers specific aspects of resource management but distills historical knowledge gathered from the myriad of AI-based decision-making methods to develop well informed AI models and eclectic management solutions. The various advances in the field of computing need to be considered in tandem to bolster AI research and build holistic AI-based methods for emerging application areas, future technologies and next-generation users.

\section*{Acknowledgments}
This work is supported by the the President’s Ph.D. Scholarship at Imperial College London and Australian Research Council Discovery Project. We thank Shikhar Tuli, Zifeng Niu, Runan Wang, Matthew Sheldon and William Plumb for helpful discussions.

% *************************

\bibliographystyle{elsarticle-num}
\bibliography{references}
% \vfill{}
\begin{wrapfigure}{l}{25mm} 
    \includegraphics[width=1in,height=1.25in,clip,keepaspectratio]{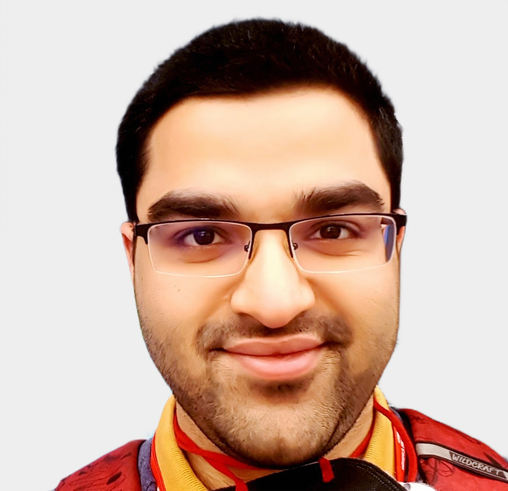}
  \end{wrapfigure}\par
  \textbf{Shreshth Tuli} is a President's Ph.D. Scholar at the Department of Computing, Imperial College London, UK. Prior to this he was an undergraduate student at the Department of Computer Science and Engineering at Indian Institute of Technology - Delhi, India. He has worked as a visiting research fellow at the CLOUDS Laboratory, School of Computing and Information Systems, the University of Melbourne, Australia. His research interests include Fog Computing and Deep Learning. For further information, visit \url{https://shreshthtuli.github.io/}.\par
% \vfill{}
\begin{wrapfigure}{l}{25mm} 
    \includegraphics[width=1in,height=1.25in,clip,keepaspectratio]{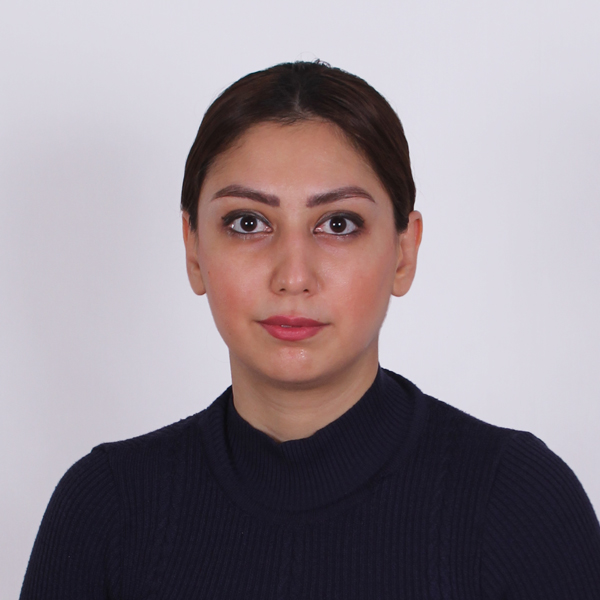}
  \end{wrapfigure}\par
  \textbf{Fatemeh Mirhakimi} is a current Master of Research student at Western Sydney University, Australia. She received B.S and M.S. in Computer Engineering from the Azad University of Arak, respectively.   She has conducted a research project about power-consuming management in hybrid vehicles based on fuzzy-genetic methods. Her research interests include cloud and edge computing, the internet of things, and reliability and fault tolerance. \par
\vfill{}
\begin{wrapfigure}{l}{25mm} 
    \includegraphics[width=1in,height=1.25in,clip,keepaspectratio]{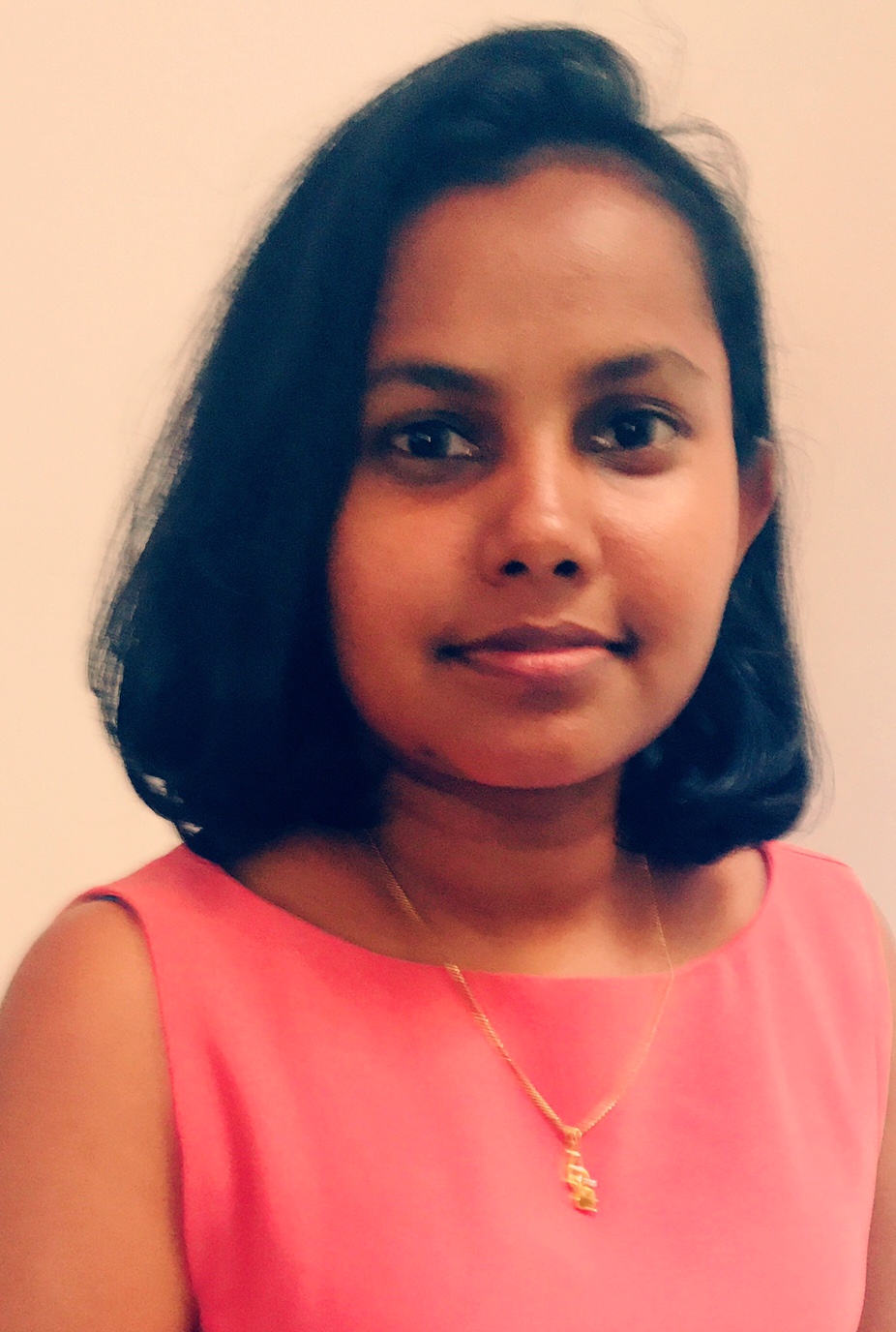}
  \end{wrapfigure}\par
  \textbf{Samodha Pallewatta} is a Ph.D. student at the Cloud Computing and Distributed Systems (CLOUDS) Laboratory, Department of Computing and Information Systems, The University of Melbourne, Australia. Her research interests encompass Fog/Edge Computing, Internet of Things (IoT), Microservice architecture and Distributed Systems. She is one of the contributors of the iFogSim simulator, used extensively for resource management research in Fog/Edge computing. \par
\vfill{}
\begin{wrapfigure}{l}{25mm} 
    \includegraphics[width=1in,height=1.25in,clip,keepaspectratio]{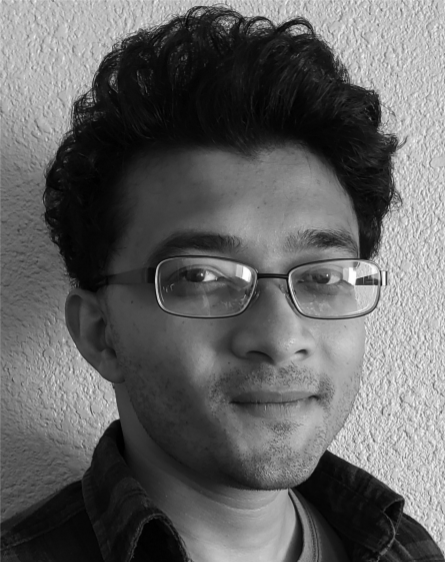}
  \end{wrapfigure}\par
  \textbf{Syed Zawad} received the B.Sc. degree in computer science and engineering from BRAC University, Dhaka, Bangladesh. He is currently pursuing the Ph.D. degree with the Department of Computer Science and Engineering, University of Nevada, Reno, NV, USA. He has interned as a Researcher with Baidu, Sunnyvale, CA, USA. He also has three years of work experience as a Software Engineer for Web applications. His research area of interest is in high performance computing, deep learning, federated learning, and neural architecture search.\par
\vfill{}
\begin{wrapfigure}{l}{25mm} 
    \includegraphics[width=1in,height=1.25in,clip,keepaspectratio]{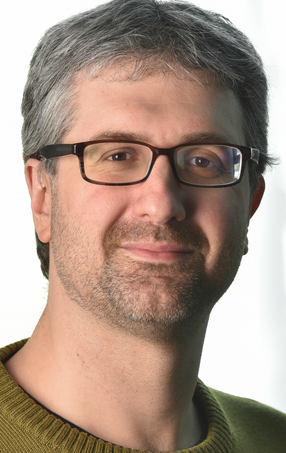}
  \end{wrapfigure}\par
  \textbf{Giuliano Casale} joined the Department of Computing at Imperial College London in 2010, where he is currently a Reader. Previously, he worked as a research scientist and consultant in the capacity planning industry. He teaches and does research in performance engineering and cloud computing, topics on which he has published more than 100 refereed papers. His research work has received multiple awards, recently the best paper award at ACM SIGMETRICS. He serves on the editorial boards of IEEE TNSM and ACM TOMPECS and as current chair of ACM SIGMETRICS.\par
\vfill{}
\begin{wrapfigure}{l}{25mm} 
    \includegraphics[width=1in,height=1.25in,clip,keepaspectratio]{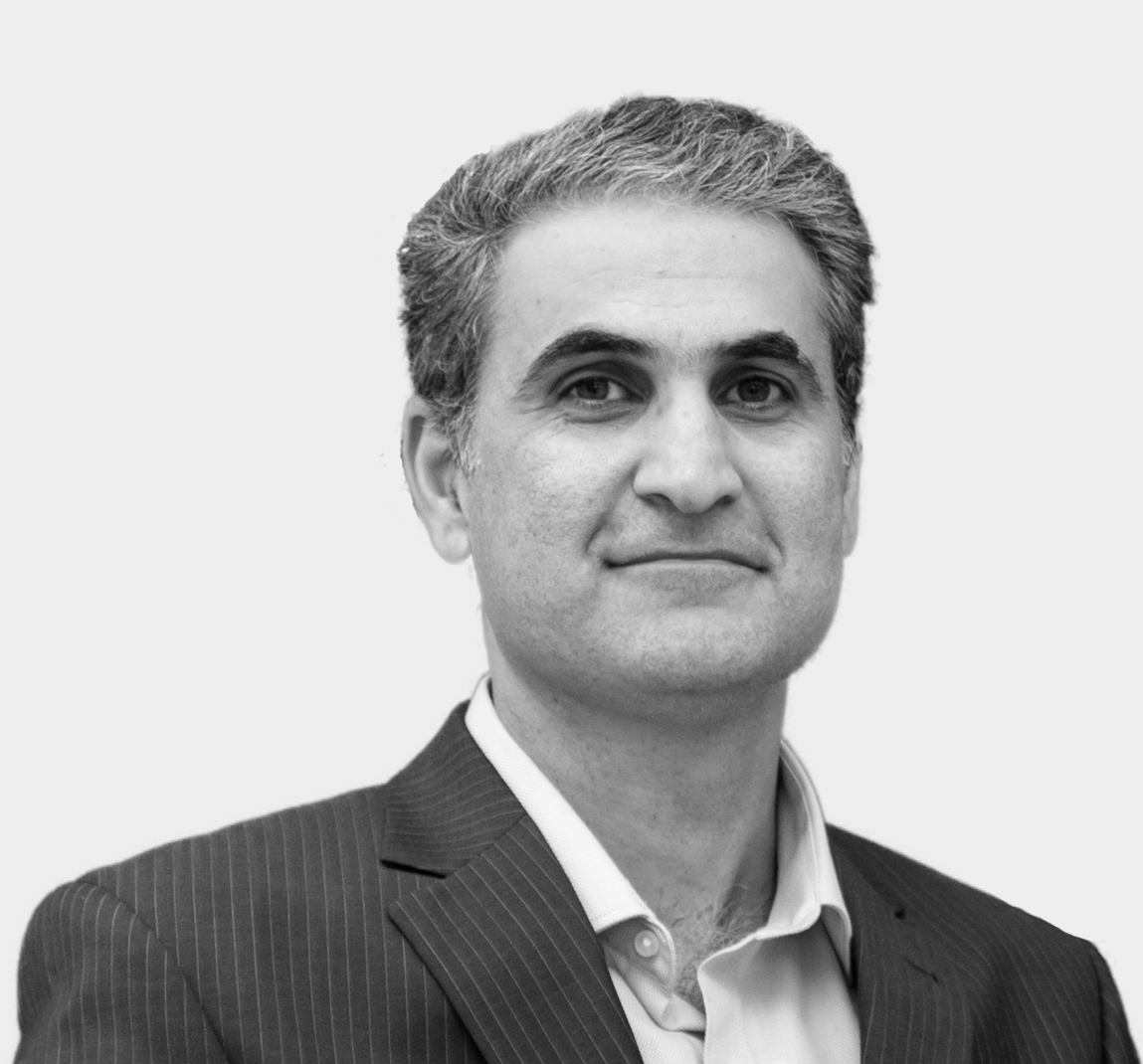}
  \end{wrapfigure}\par
  \textbf{Bahman Javadi} is an Associate Professor in Networking and Cloud Computing at Western Sydney University, Australia. Prior to this appointment, he was a Research Fellow at the University of Melbourne and a Postdoctoral Fellow at the INRIA Rhone-Alpes, France. His research interests include Cloud computing, Edge computing, performance evaluation of large-scale distributed computing systems, and reliability and fault tolerance. He is a Senior Member of ACM, Senior Member of IEEE and Senior Fellow of Advance HE of UK. His website is: \url{https://staff.cdms.westernsydney.edu.au/~bjavadi/}. \par
\vfill{}
\begin{wrapfigure}{l}{25mm} 
    \includegraphics[width=1in,height=1.25in,clip,keepaspectratio]{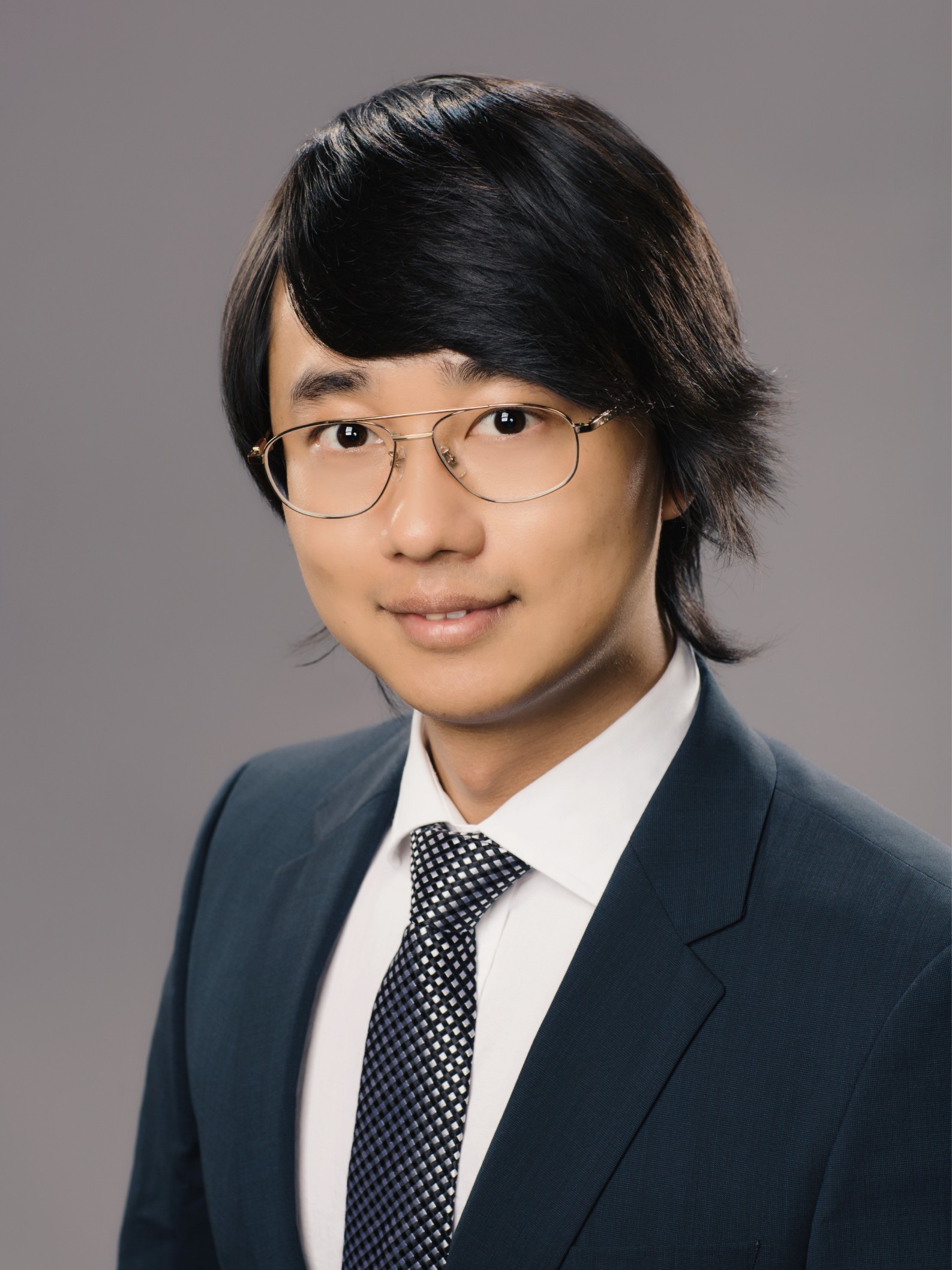}
  \end{wrapfigure}\par
  \textbf{Feng Yan} is an Assistant Professor of Computer Science and Engineering at University of Nevada, Reno and the director of the Intelligent Data and Systems Lab. He received M.S. and Ph.D. degrees in Computer Science from the College of William and Mary and worked at Microsoft Research and HP Labs. His research bridges the fields of big data, machine learning, and systems. He is a recipient of the Best Student Paper Award of IEEE CLOUD 2018, the Best Paper Award of CLOUD 2019, and the Best Student Paper Award of ITNG 2021, as well as the NSF CAREER Award, the NSF CRII Award, and the Regents' Rising Researcher Award. \par
% \vfill{}
\begin{wrapfigure}{l}{25mm} 
    \includegraphics[width=1in,height=1.25in,clip,keepaspectratio]{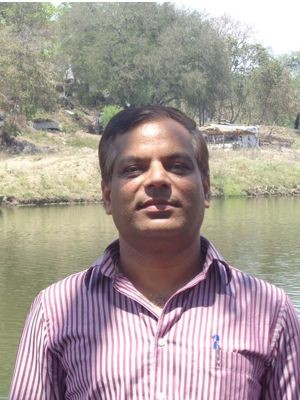}
  \end{wrapfigure}\par
  \textbf{Rajkumar Buyya} is a Redmond Barry Distinguished Professor and Director of the Cloud Computing and Distributed Systems (CLOUDS) Laboratory at the University of Melbourne, Australia. He has authored over 625 publications and seven text books including "Mastering Cloud Computing" published by McGraw Hill, China Machine Press, and Morgan Kaufmann for Indian, Chinese and international markets respectively.  He is one of the highly cited authors in computer science and software engineering worldwide (h-index=154, g-index=322, 124,800+ citations).  \par
% \vfill{}
\begin{wrapfigure}{l}{25mm} 
    \includegraphics[width=1in,height=1.25in,clip,keepaspectratio]{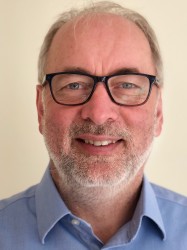}
  \end{wrapfigure}\par
  \textbf{Nicholas R. Jennings} is the Vice-Chancellor and President of Loughborough University. He is an internationally-recognised authority in the areas of AI, autonomous systems, cyber-security and agent-based computing. He is a member of the UK government’s AI Council, the governing body of the Engineering and Physical Sciences Research Council, and chair of the Royal Academy of Engineering’s Policy Committee.  Before Loughborough, he was the Vice-Provost for Research and Enterprise and Professor of Artificial Intelligence at Imperial College London, the UK's first Regius Professor of Computer Science (a post bestowed by the monarch to recognise exceptionally high quality research) and the UK Government’s first Chief Scientific Advisor for National Security.\par
\vfill{}

\end{document}